\documentclass[lettersize,journal]{IEEEtran}
\usepackage{amsmath,amsfonts}
\usepackage{cases}
\usepackage[ruled]{algorithm2e} 
\usepackage{array}
\usepackage{subfigure}
\usepackage{textcomp}
\usepackage{stfloats}
\usepackage{url}
\usepackage{verbatim}
\usepackage{graphicx}
\usepackage{cite}
\usepackage{amssymb}
\newtheorem{remark}{Remark}
\usepackage{xspace}
\usepackage{float}
\newtheorem{proof}{Proof}
\begin{document}

\title{Quantized criterion-based kernel recursive least squares adaptive filtering for time series prediction
}

\author{Jiacheng He, Gang Wang, Kun Zhang, Shan Zhong, Bei Peng
\thanks{This paper was produced by the IEEE Publication Technology Group. They are in Piscataway, NJ.}
\thanks{Manuscript received April 19, 2021; revised August 16, 2021.}}

\markboth{Journal of \LaTeX\ Class Files,~Vol.~14, No.~8, August~2021}%
{Shell \MakeLowercase{\textit{et al.}}: A Sample Article Using IEEEtran.cls for IEEE Journals}


\maketitle

\begin{abstract}
The robustness of the Kernel Recursive Least Squares (KRLS) algorithm has recently seen improvements through its integration with more robust information-theoretic learning criteria, such as Minimum Error Entropy (MEE) and Generalized MEE (GMEE). This integration has also led to enhancements in the computational efficiency of KRLS-type algorithms. To alleviate the computational burden associated with KRLS-type algorithms, this paper introduces the quantized GMEE (QGMEE) criterion. This criterion is integrated with the KRLS algorithm, giving rise to two novel KRLS-type algorithms: Quantized Kernel Recursive MEE (QKRMEE) and Quantized Kernel Recursive GMEE (QKRGMEE). Furthermore, the paper thoroughly investigates the mean error behavior, the mean square error behavior, and the computational complexity of the proposed algorithms. To validate their effectiveness, both simulation studies and experiments with real-world data are conducted. 
\end{abstract}


\section{Introduction}\label{Introduction}
Time series prediction \cite{10209568} is a critically important data analysis technique with the primary goal of predicting future trends, thereby enabling informed decision-making by professionals. This technique has been widely applied in many fields, such as finance \cite{cao2019financial}, meteorology \cite{faisal2022neural}, and disease outbreak prediction \cite{wang2020time}, and others \cite{10075373,9701926,8742529}. The method based on kernel adaptive filtering has been widely applied in time series prediction, such as kernel affine projection algorithms (KAPAs) \cite{liu2008kernel}, kernel least mean square (KLMS) \cite{zhao2021adaptive}, kernel recursive least squares (KRLS) \cite{guo2020improved}, extended kernel recursive least squares (EX-KRLS) \cite{4907077}, and other similar algorithms exemplify typical kernel adaptive filtering (KAF) techniques \cite{MA2017101}.

The KRLS algorithm is particularly renowned for its consistent performance in nonlinear systems when Gaussian noise is present. However, non-Gaussian noise frequently poses challenges in various practical applications, including underwater communications \cite{HE2022}, parameter identification \cite{7862211,8345754}, and acoustic echo cancellation \cite{HE2023108787,HE20221362}. The existence of non-Gaussian noise notably deteriorates the performance of the KRLS algorithm, especially when it is evaluated using the mean suare error (MSE) criterion.

Hence, novel adaptation criteria rooted in information-theoretic learning (ITL) \cite{liu2009information} are introduced to surmount these challenges. These criteria harness higher-order statistics of distributions. Noteworthy examples of ITL criteria frequently used to improve the learning efficiency of algorithms include the maximum correntropy criterion (MCC) \cite{liu2007correntropy}, the generalized maximum correntropy criterion (GMCC) \cite{chen2016generalized,ZHAO202266} and the minimum error entropy (MEE) criterion \cite{10.5555/1855180,10239447}. Leveraging these learning criteria, the KLMS and KRLS algorithms are synergized with the MCC, leading to the development of the kernel maximum correntropy (KMC) algorithm \cite{6033473,8345754} and kernel recursive maximum correntropy (KRMC) algorithms \cite{wu2015kernel,WANG2019424}, respectively. For further enhancing the performance of KAF algorithms, the GMCC is integrated with KAF algorithms, giving rise to the generalized kernel maximum correntropy (GKMC) algorithm \cite{7727409} and kernel recursive generalized maximum correntropy (KRGMCC) algorithm \cite{8064727}. The MEE criterion is recognized to outperform the MCC \cite{chen2019minimum}, leading to the exploration of the kernel MEE (KMEE) algorithm \cite{CHEN2013160} and the kernel recursive MEE (KRMEE) algorithm \cite{WANG2022108410}. Although the Gaussian kernel function is commonly selected for the original MEE due to its smoothness and strict positive definiteness, it might not be universally optimal \cite{chen2016generalized}. Introduction of the generalized Gaussian density prompts the formulation of the generalized minimum error entropy (GMEE) criterion \cite{HE2023109188}. Moreover, the kernel recursive GMEE (KRGMEE) algorithm, exhibiting superior performance, is also derived in \cite{HE2023109188}. However, it is important to note that due to the double summation of GMEE criteria, the computational complexity of the information potential (IP), the fundamental cost associated with GMEE in ITL, becomes quadratic concerning the sample count. Incorporation of GMEE heightens the efficacy of the KRLS algorithm, while introducing an increase in computational complexity, particularly for extensive datasets.

In this study, we employ a quantizer \cite{chen2018quantized} to mitigate the computational complexity of the IP associated with both the GMEE and MEE criteria. By integrating the quantized generalized minimum error entropy (QGMEE) criterion with the kernel recursive least squares (KRLS) method, we introduce two novel algorithms: the quantized kernel recursive GMEE (QKRGMEE) algorithm and the quantized kernel recursive MEE (QKRMEE) algorithm. Notably, the QKRMEE algorithm is a specific variant derived from the QKRGMEE algorithm. Furthermore, we delve into several properties of the QGMEE criterion to enhance the theoretical foundation of QGMEE. We then present the mean error behavior and mean square error behavior of the proposed quantized algorithm. Additionally, we conduct a comparative analysis of the performance and computational complexity of the introduced quantized methodologies against the KRGMEE and KRGMEE algorithms.

The main contributions of this study are the following.

(1) The properties of the proposed QGMEE criterion are analyzed and discussed.

(2) Two quantized kernel recursive least squares algorithms (QKRMEE and QKRGMEE), with lower computational complexity, are proposed.

(3) The performance of the proposed algorithm is verified using Electroencephalogram (EEG) data.

The remainder of the study is organized as follows. The properties of the QGMEE are shown in Section \ref{secproqgmee}. The proposed KRLS algorithms are presented in Section \ref{qgmeeprpose}. Section \ref{perfroanaly} and Section \ref{simulation} are performance analyses and simulations, respectively. Finally, Section \ref{conclusion} provides the conclusion.

\section{The properties of the quantized generalized error entropy}\label{secproqgmee}
\subsection{Quantized generalized error entropy}
From \cite{HE2023109188},the definition of the quantized generalized error entropy is
\begin{align}
{H_\mu }\left( e \right) = \frac{1}{{1 - \mu }}\log V_{\alpha ,\beta }^\mu \left( e \right),
\end{align}
where ${\mu \left( {\mu  \ne 1,\mu  > 0} \right)}$ stands for the entropy order and ${e}$ for the error between ${X}$ and ${Y}$. A continuous variable IP $V_{\alpha ,\beta }^\mu \left( e \right)$ can be written as 
\begin{equation}\label{zhkuxikueajyibalz}
\begin{split}
V_{\alpha ,\beta }^\mu \left( {{\boldsymbol{X}},{\boldsymbol{Y}}} \right){\text{  =  }}V_{\alpha ,\beta }^\mu \left( e \right) = \int {p_{\alpha ,\beta }^\mu \left( e \right)de}  = \operatorname{E} \left[ {p_{\alpha ,\beta }^{\mu  - 1}\left( e \right)} \right],
\end{split}
\end{equation}
where $\operatorname{E} \left[  \cdot  \right]$ represents the expectation operation, and $\mu$ is set to 2 in \cite{HE2023109188}.
${{p_{\alpha ,\beta }}\left(  \cdot  \right)}$ denotes the PDF of error ${e}$, as well as ${{p_{\alpha ,\beta }}\left(  \cdot  \right)}$ can be estimated using
\begin{equation}
\begin{split}\label{eikjxkbetalphg}
{p_{\alpha ,\beta }}\left( e \right) \approx {\hat p_{\alpha ,\beta }}\left( e \right) = \frac{1}{L}\sum\limits_{i = 1}^L {{G_{\alpha ,\beta }}\left( {e - {e_i}} \right)} .
\end{split}
\end{equation}

Only a limited number of error sets ${\left\{ {{e_i}} \right\}_{i = 1}^{L}}$ may be obtained in actual applications, and $L$ is the length of Parzen window. Substituting (\ref{eikjxkbetalphg}) into (\ref{zhkuxikueajyibalz}) with ${\mu {\text{ = }}2}$, and one method for estimating the IP ${{V_{\alpha ,\beta }}\left( {{\boldsymbol{X}},{\boldsymbol{Y}}} \right)}$ is
\begin{equation}
\begin{split}\label{ValphabetamaoLFONE}
\begin{gathered}
  {{\hat V}_{\alpha ,\beta }}\left( {{\boldsymbol{X}},{\boldsymbol{Y}}} \right) = {{\hat V}_{\alpha ,\beta }}\left( {\boldsymbol{e}} \right) =  \hfill \\
  \frac{1}{L}\sum\limits_{i = 1}^L {{{\hat p}_{\alpha ,\beta }}} \left( {{e_i}} \right) = \frac{1}{{{L^2}}}\sum\limits_{i = 1}^L {\sum\limits_{j = 1}^L {{G_{\alpha ,\beta }}} } \left( {{e_i} - {e_j}} \right), \hfill \\ 
\end{gathered} 
\end{split}
\end{equation}
with
\begin{align}\label{gaussalphabeta}
{G_{\alpha ,\beta }}\left( e \right) = \frac{\alpha }{{2\beta \Gamma \left( {1/\alpha } \right)}}\exp \left( { - \frac{{{{\left| e \right|}^\alpha }}}{{{\beta ^\alpha }}}} \right),
\end{align}
where ${G_{\alpha ,\beta }}\left(  \cdot  \right)$ the generalized Gaussian density function \cite{varanasi1989parametric}, $\alpha$ and $\beta  > 0$ refer to the shape and scale parameters, $\left|  \cdot  \right|$ is taking the absolute value, and $\Gamma \left(  \cdot  \right)$ stands for the gamma function. Using estimator \eqref{ValphabetamaoLFONE}, the quadratic IP may be calculated.

A quantizer ${\text{Q}}\left( {{e_i},\gamma } \right) \in C$ \cite{chen2018quantized} (quantization threshold $\gamma$ is employed to obtain a codebook $C = \left\{ {{c_1},{c_2}, \cdots {c_H} \in {\mathbb{R}^1}} \right\}$ in order to lessen the computational load on GMEE, where $H$ represents the number of quantized sample errors. The empirical information potential can be simplified to
\begin{align}\label{palbeeiqlf1}
\begin{gathered}
  {{\hat V}_{\alpha ,\beta }}\left( {\boldsymbol{e}} \right) = \frac{1}{L}\sum\limits_{i = 1}^L {{{\hat p}_{\alpha ,\beta }}} \left( {{e_i}} \right) \approx \hat V_{\alpha ,\beta }^Q\left( {\boldsymbol{e}} \right) \hfill \\
   = \frac{1}{{{L^2}}}\sum\limits_{i = 1}^L {\sum\limits_{j = 1}^L {{G_{\alpha ,\beta }}} } \left[ {{e_i} - {\text{Q}}\left[ {{e_j},\gamma } \right]} \right] \hfill \\
   = \frac{1}{{{L^2}}}\sum\limits_{i = 1}^L {\sum\limits_{h = 1}^H {{H_h}{G_{\alpha ,\beta }}} } \left[ {{e_i} - {c_h}} \right] \hfill \\
   = \frac{1}{L}\hat p_{\alpha ,\beta }^Q\left( {{e_i}} \right), \hfill \\ 
\end{gathered}
\end{align}
where ${{H_h}}$ is the number of quantized error samples $c_h$. And, one can get ${L = \sum\nolimits_{h = 1}^H {{H_h}} }$ and ${\int {\hat p_{\alpha ,\beta }^Q} \left( e \right)de = 1}$. The adjustable threshold $\gamma$ controls the number of elements in the codebook and thus the computational effort of the algorithm.

When $\alpha {\text{ = 2}}$, the QGMEE criterion translates into QMEE criterion \cite{chen2018quantized} with the following form:
\begin{align}
\begin{gathered}
  {{\hat V}_\sigma }\left( {\boldsymbol{e}} \right) = \frac{1}{L}\sum\limits_{i = 1}^L {{{\hat p}_\sigma }} \left( {{e_i}} \right) \approx \hat V_\sigma ^Q\left( {\boldsymbol{e}} \right) \hfill \\
   = \frac{1}{{{L^2}}}\sum\limits_{i = 1}^L {\sum\limits_{h = 1}^H {{H_h}{G_\sigma }\left[ {{e_i} - {c_h}} \right]} } , \hfill \\ 
\end{gathered}  
\end{align}
where ${G_\sigma }\left(  \cdot  \right)$ is Gaussian function with the form of ${G_\sigma }\left( e \right) = {1 \mathord{\left/
 {\vphantom {1 {\sqrt {2\pi } }}} \right.
 \kern-\nulldelimiterspace} {\sqrt {2\pi } }}\sigma \exp \left[ { - \left( {{1 \mathord{\left/
 {\vphantom {1 {2{\sigma ^2}}}} \right.
 \kern-\nulldelimiterspace} {2{\sigma ^2}}}} \right){e^2}} \right]$, parameter $\sigma$ represents the bandwidth.

From another point of view, a quantizer is a clustering algorithm that classifies the set of errors at a certain distance and the distance is the quantization threshold $\gamma$. Predictably, the larger the gamma, the fewer the number $H$ of classes the error set is divided into.
 
\subsection{Properties}
\emph{Property 1:} When $\gamma=0$, one can be obtain that ${\hat V_{\alpha ,\beta }}\left( {\boldsymbol{e}} \right) = \hat V_{\alpha ,\beta }^Q\left( {\boldsymbol{e}} \right).$
\begin{proof}
In the case of $\gamma=0$, the code book is $C = \left\{ {{e_1},{e_2}, \cdots ,{e_L}} \right\}$. According to \eqref{palbeeiqlf1}, one can obtain ${\hat V_{\alpha ,\beta }}\left( {\boldsymbol{e}} \right) = \hat V_{\alpha ,\beta }^Q\left( {\boldsymbol{e}} \right).$
\end{proof}
\emph{Property 2:} The proposed cost function $\hat V_{\alpha ,\beta }^Q\left( {\boldsymbol{e}} \right)$ is bounded, which can be expressed specifically as $\hat V_{\alpha ,\beta }^Q\left( {\boldsymbol{e}} \right) \leqslant {\alpha  \mathord{\left/
 {\vphantom {\alpha  {2\beta \Gamma \left( {1/\alpha } \right)}}} \right.
 \kern-\nulldelimiterspace} {2\beta \Gamma \left( {1/\alpha } \right)}}$, with equality if and only if ${e_1} = {e_2} =  \cdots  = {e_L}$.
\begin{proof}
From \eqref{palbeeiqlf1} and \eqref{gaussalphabeta}, we can obtain
\begin{align}
\hat V_{\alpha ,\beta }^Q\left( {\boldsymbol{e}} \right) = \frac{1}{{{L^2}}}\sum\limits_{i = 1}^L {\sum\limits_{j = 1}^L {\frac{\alpha }{{2\beta \Gamma \left( {1/\alpha } \right)}}\exp \left( { - \frac{{\left| {{e_i} - {\text{Q}}\left( {{e_j},\gamma } \right)} \right|}}{{{\beta ^\alpha }}}} \right)} } ,
\end{align}
since ${G_{\alpha ,\beta }}\left( e \right) \leqslant {\alpha  \mathord{\left/
 {\vphantom {\alpha  {2\beta \Gamma \left( {1/\alpha } \right)}}} \right.
 \kern-\nulldelimiterspace} {2\beta \Gamma \left( {1/\alpha } \right)}}$ with equality if and only if $e = 0$. Therefore, one can obtain
 \begin{align}
\hat V_{\alpha ,\beta }^Q\left( {\boldsymbol{e}} \right) \leqslant \frac{1}{{{L^2}}}\sum\limits_{i = 1}^L {\sum\limits_{j = 1}^L {\frac{\alpha }{{2\beta \Gamma \left( {1/\alpha } \right)}} = } } \frac{\alpha }{{2\beta \Gamma \left( {1/\alpha } \right)}}.
 \end{align}
\end{proof}

\emph{Property 3:}
It holds that $\hat V_{\alpha ,\beta }^Q\left( {\boldsymbol{e}} \right) = \sum\nolimits_{h = 1}^H {{a_h}\hat p\left( {{c_h}} \right)} $, where ${a_h} = {{{H_h}} \mathord{\left/
 {\vphantom {{{H_h}} L}} \right.
 \kern-\nulldelimiterspace} L}$, and one can obtain $\sum\limits_{h = 1}^H {{a_h}}  = 1$.
\begin{proof}
It can easily be deduced that
\begin{align}\label{pchmaoahxih}
\begin{gathered}
  \hat V_{\alpha ,\beta }^Q\left( {\boldsymbol{e}} \right) = \sum\limits_{h = 1}^H {\frac{{{H_h}}}{L}\left( {\frac{1}{L}\sum\limits_{i = 1}^L {{G_{\alpha ,\beta }}\left( {{e_i} - {c_h}} \right)} } \right)}  \hfill \\
   = \sum\limits_{h = 1}^H {{a_h}\hat p\left( {{c_h}} \right)} . \hfill \\ 
\end{gathered} 
\end{align}
\end{proof}

\begin{remark}
From property 3, one can obtain that quantized IP is the weighted sum of Parzen's PDF estimator, and this weight is determined by the number $H_h$ in class $h$.  
\end{remark}

\emph{Property 4:} The generalized correntropy criterion (GCC) is a special case of the QGMEE criterion.
\begin{proof}
When $\gamma$ is large enough so that $H=1$, we can obtain $\hat V_{\alpha ,\beta }^Q\left( {\boldsymbol{e}} \right) = \hat p\left( {{c_h}} \right)$ from \eqref{pchmaoahxih}. When $H=1$ and $C = \left\{ 0 \right\}$, for the more special case, \eqref{pchmaoahxih} can be further written as
\begin{align}
\hat V_{\alpha ,\beta }^Q\left( {\boldsymbol{e}} \right) = \frac{1}{L}\sum\limits_{i = 1}^L {{G_{\alpha ,\beta }}\left( {{e_i}} \right)},
\end{align}
\end{proof}
which is GCC in \cite{chen2016generalized}.
\begin{remark}
The GCC measures the local similarity at zero, while the QGMEE criterion measures the average similarity about every $c_h$.
\end{remark}

\emph{Property 5:} When scale parameter ${\beta}$ is sufficiently big, we can get
\begin{align}
\begin{gathered}
  \hat V_{\alpha ,\beta }^Q\left( {\boldsymbol{e}} \right) \approx  \hfill \\
  \frac{\alpha }{{2\beta \Gamma \left( {1/\alpha } \right)}} - \frac{\alpha }{{2{{\left| \beta  \right|}^{\alpha  + 1}}\Gamma \left( {1/\alpha } \right)}}\sum\limits_{i = 1}^L {\sum\limits_{h = 1}^H {\frac{{{H_h}}}{{{L^2}}}{{\left| {{e_i} - {c_h}} \right|}^\alpha }} } , \hfill \\ 
\end{gathered} 
\end{align}
where ${1 \mathord{\left/
 {\vphantom {1 L}} \right.
 \kern-\nulldelimiterspace} L}\sum\nolimits_{i = 1}^L {{{\left| {{e_i} - {c_h}} \right|}^\alpha }} $ is the $\alpha$-order moment of error about $c_h$

\begin{proof}
By using Taylor series, \eqref{palbeeiqlf1} can be rewritten as
\begin{align}
\begin{gathered}
  \hat V_{\alpha ,\beta }^Q\left( {\boldsymbol{e}} \right) = \frac{1}{{{L^2}}}\sum\limits_{i = 1}^L {\sum\limits_{h = 1}^H {{H_h}{G_{\alpha ,\beta }}\left( {{e_i} - {c_h}} \right)} }  \hfill \\
   = \frac{\alpha }{{2\beta \Gamma \left( {1/\alpha } \right){L^2}}}\sum\limits_{i = 1}^L {\sum\limits_{h = 1}^H {\sum\limits_{t = 0}^\infty  {\frac{1}{{t!}}{H_h}{{\left( { - {{\left| {\frac{{{e_i} - {c_h}}}{\beta }} \right|}^\alpha }} \right)}^t}} } } , \hfill \\ 
\end{gathered} 
\end{align}
and when $\beta$ is sufficiently big, we can obtain
\begin{align}
\begin{gathered}
  \hat V_{\alpha ,\beta }^Q\left( {\boldsymbol{e}} \right) \approx \frac{\alpha }{{2\beta \Gamma \left( {1/\alpha } \right){L^2}}}\sum\limits_{i = 1}^L {\sum\limits_{h = 1}^H {{H_h}\left[ {1 - \frac{1}{{{{\left| \beta  \right|}^\alpha }}}{{\left| {{e_i} - {c_h}} \right|}^\alpha }} \right]} }  \hfill \\
   = \frac{\alpha }{{2\beta \Gamma \left( {1/\alpha } \right)}} - \frac{\alpha }{{2{{\left| \beta  \right|}^{\alpha  + 1}}\Gamma \left( {1/\alpha } \right)}}\sum\limits_{i = 1}^L {\sum\limits_{h = 1}^H {\frac{{{H_h}}}{{{L^2}}}{{\left| {{e_i} - {c_h}} \right|}^\alpha }} } . \hfill \\ 
\end{gathered} 
\end{align}
\end{proof}

\emph{Property 6:} In the case of regression model ${\boldsymbol{f}}\left( {\boldsymbol{u}} \right) = {{\boldsymbol{w}}^T}{\boldsymbol{u}}$ with a vector ${\boldsymbol{w}}$ of weights that need to be estimated, and ${\boldsymbol{w}}$ the optimal solution based on the QGMEE criterion is ${\boldsymbol{w}} = N_{QGMEE}^{ - 1}{M_{QGMEE}}$, where ${M_{QGMEE}} = \sum\nolimits_{i = 1}^L {\sum\nolimits_{h = 1}^H {{H_h}{{\text{G}}_{\alpha ,\beta }}\left( {{e_i} - {c_h}} \right){{\left| {{e_i} - {c_h}} \right|}^{\alpha  - 2}}\left( {{d_i} - {c_h}} \right){{\boldsymbol{u}}_i}} } $ and ${N_{QGMEE}} = \sum\nolimits_{i = 1}^L {\sum\nolimits_{h = 1}^H {{H_h}{{\text{G}}_{\alpha ,\beta }}\left( {{e_i} - {c_h}} \right){{\left| {{e_i} - {c_h}} \right|}^{\alpha  - 2}}{{\boldsymbol{u}}_i}{\boldsymbol{u}}_i^T} } $.
\begin{proof}
The gradient of the cost function ${\hat V_{\alpha ,\beta }^Q\left( {\boldsymbol{e}} \right)}$ with respect to ${\boldsymbol{w}}$ can be written as
\begin{align}\label{gonghsi15wfiwu}
\begin{gathered}
  \frac{{\partial \hat V_{\alpha ,\beta }^Q\left( {\boldsymbol{e}} \right)}}{{\partial {\boldsymbol{w}}}} \hfill \\
   = \frac{1}{{{L^2}}}\frac{\alpha }{{{\beta ^\alpha }}}\sum\limits_{i = 1}^L {\sum\limits_{h = 1}^H {\left[ \begin{gathered}
  {H_h}{{\text{G}}_{\alpha ,\beta }}\left( {{e_i} - {c_h}} \right){\left| {{e_i} - {c_h}} \right|^{\alpha  - 1}} \hfill \\
   \times \operatorname{sign} \left( {{e_i} - {c_h}} \right){{\boldsymbol{u}}_i} \hfill \\ 
\end{gathered}  \right]} }  \hfill \\
   = \frac{\alpha }{{{L^2}{\beta ^\alpha }}}\sum\limits_{i = 1}^L {\sum\limits_{h = 1}^H {\left[ \begin{gathered}
  {H_h}{{\text{G}}_{\alpha ,\beta }}\left( {{e_i} - {c_h}} \right){\left| {{e_i} - {c_h}} \right|^{\alpha  - 2}} \hfill \\
   \times \left[ {\left( {{d_i} - {c_h}} \right) - {\boldsymbol{w}}_n^T{\varphi _i}} \right]{{\boldsymbol{u}}_i} \hfill \\ 
\end{gathered}  \right]} }  \hfill \\
   = \frac{\alpha }{{{L^2}{\beta ^\alpha }}}\sum\limits_{i = 1}^L {\sum\limits_{h = 1}^H {{H_h}{{\text{G}}_{\alpha ,\beta }}\left( {{e_i} - {c_h}} \right){{\left| {{e_i} - {c_h}} \right|}^{\alpha  - 2}}\left( {{d_i} - {c_h}} \right){{\boldsymbol{u}}_i}} }  \hfill \\
   - \frac{\alpha }{{{L^2}{\beta ^\alpha }}}\sum\limits_{i = 1}^L {\sum\limits_{h = 1}^H {{H_h}{{\text{G}}_{\alpha ,\beta }}\left( {{e_i} - {c_h}} \right){{\left| {{e_i} - {c_h}} \right|}^{\alpha  - 2}}{{\boldsymbol{u}}_i}{\boldsymbol{u}}_i^T{\boldsymbol{w}}} } . \hfill \\
   = \frac{\alpha }{{{L^2}{\beta ^\alpha }}}{M_{QGMEE}} - \frac{\alpha }{{{L^2}{\beta ^\alpha }}}{N_{QGMEE}}{\boldsymbol{w}}. \hfill \\ 
\end{gathered} 
\end{align}
\end{proof}
 Setting \eqref{gonghsi15wfiwu} equal to zero, and one can obtain ${\boldsymbol{w}} = N_{QGMEE}^{ - 1}{M_{QGMEE}}$, which proves the property 6. Property 6 is utilized to deal with regression problem in \cite{HE2023109188}, and the QGMEE adaptive filtering is developed.

\section{Kernel adaptive filtering based on QGMEE}\label{qgmeeprpose}

The input vector ${{\boldsymbol{u}}_n} \in \mathbb{U}$ is considered to be transformed into a hypothesis space $\mathbb{K}$ by the nonlinear mapping ${\boldsymbol{f}}\left(  \cdot  \right)$. The input space $\mathbb{U}$ is a compact domain of ${\mathbb{R}^M}$, and the output ${d_n} \in {\mathbb{R}^1}$ can be described from
\begin{align}
{d_n} = {\boldsymbol{f}}\left( {{{\boldsymbol{u}}_n}} \right) + {v_n},
\end{align}
where ${v_n}$ denote the zero-mean noise. RKHS with a Mercer kernel $\kappa \left( {{\boldsymbol{X}},{\boldsymbol{Y}}} \right)$ will be the learning hypotheses space. The norms used in this study are all ${l_2}$-norms. The commonly used Gaussian kernel with bandwidth $\sigma$ is utilized:
\begin{align}\label{kernelxy}
\kappa \left( {{\boldsymbol{X}},{\boldsymbol{Y}}} \right) = {G_\sigma }\left( {{\boldsymbol{X}} - {\boldsymbol{Y}}} \right) = \frac{1}{{\sqrt {2\pi } \sigma }}\exp \left( { - \frac{1}{{2{\sigma ^2}}}{{\left\| {{\boldsymbol{X}} - {\boldsymbol{Y}}} \right\|}^2}} \right).
\end{align}
Theoretically, every Mercer kernel will produce a distinct hypothesis feature space \cite{smola2002support}. As a result, the input data $\left\{ {{{\boldsymbol{u}}_1},{{\boldsymbol{u}}_2}, \cdots {{\boldsymbol{u}}_N}} \right\}$ will be translated into the feature space as $\left\{ {{{\boldsymbol{\varphi }}_1},{{\boldsymbol{\varphi }}_2}, \cdots {{\boldsymbol{\varphi }}_N}} \right\}$ ($N$ represents the total number of the input data), rendering it impossible to do a straight calculation. Instead, using the well-known "kernel trick" \cite{smola2002support}, one can get the inner production from \eqref{kernelxy}:
\begin{align}
\varphi _i^T{\varphi _j} = \kappa \left( {{{\boldsymbol{u}}_i},{{\boldsymbol{u}}_j}} \right) = \frac{1}{{\sqrt {2\pi } \sigma }}\exp \left( { - \frac{1}{{2{\sigma ^2}}}{{\left\| {{{\boldsymbol{u}}_i} - {{\boldsymbol{u}}_j}} \right\|}^2}} \right).
\end{align}
The filter output can be expressed as ${\boldsymbol{w}}_n^T{{\boldsymbol{\varphi }}_n}$ for each time point $n$, where ${{\boldsymbol{w}}_n}$ is a weight vector in the high-dimensional hypothesis space $\mathbb{K}$, therefore, the output error can be written separately as
\begin{align}
{e_n} = {d_n} - {\boldsymbol{w}}_n^T{{\boldsymbol{\varphi }}_n}.
\end{align}

\subsection{Kernel recursive QGMEE and QMEE algorithm}
According to the QGMEE criterion, one can obtain the cost function with the following form:
\begin{align}\label{JQGMEErls}
\begin{gathered}
  {J_{QGMEE}}\left( {{{\boldsymbol{w}}_n}} \right) =  \hfill \\
  \frac{1}{{{L^2}}}{\lambda ^{i + h}}\sum\limits_{i = 1}^L {\sum\limits_{h = 1}^H {{H_h}{{\text{G}}_{\alpha ,\beta }}\left( {{e_i} - {c_h}} \right)} }  - \frac{1}{2}{\vartheta _1}{\left\| {{{\boldsymbol{w}}_n}} \right\|^2}, \hfill \\ 
\end{gathered} 
\end{align}
where $0 < \lambda  \leqslant 1$ stands for the exponential forgetting factor. The gradient of \eqref{JQGMEErls} with respect to ${{{\boldsymbol{w}}_n}}$ is
\begin{align}\label{respwnsmji1}
\begin{gathered}
  \frac{{\partial {J_{QGMEE}}\left( {{{\boldsymbol{w}}_n}} \right)}}{{\partial {{\boldsymbol{w}}_n}}} =  - {\vartheta _1}{{\boldsymbol{w}}_n} \hfill \\
  {\text{ + }}\frac{\alpha }{{{L^2}{\beta ^\alpha }}}\sum\limits_{i = 1}^L {\sum\limits_{h = 1}^H {\left( \begin{gathered}
  {\lambda ^{i + h}}{H_h}{{\text{G}}_{\alpha ,\beta }}\left( {{e_i} - {c_h}} \right){\left| {{e_i} - {c_h}} \right|^{\alpha  - 2}} \hfill \\
   \times \left[ {\left( {{d_i} - {c_h}} \right) - {\boldsymbol{w}}_n^T{\varphi _i}} \right]{\varphi _i} \hfill \\ 
\end{gathered}  \right)} } . \hfill \\ 
\end{gathered} 
\end{align}
By applying a formal transformation to equation \eqref{respwnsmji1}, \eqref{respwnsmji1} can be further written as
\begin{align}
\begin{gathered}
  \frac{{\partial {J_{QGMEE}}\left( {{{\boldsymbol{w}}_n}} \right)}}{{\partial {{\boldsymbol{w}}_n}}} =  \hfill \\
  \frac{\alpha }{{{L^2}{\beta ^\alpha }}}{{\boldsymbol{\Phi }}_L}{{\boldsymbol{\Lambda }}_L}{{\boldsymbol{d}}_L} - \frac{\alpha }{{{L^2}{\beta ^\alpha }}}{{\boldsymbol{\Phi }}_L}{{\boldsymbol{\Lambda }}_L}{\boldsymbol{\Phi }}_L^T{{\boldsymbol{w}}_n} - {\vartheta _1}{{\boldsymbol{w}}_n}, \hfill \\ 
\end{gathered} 
\end{align}
where 
\begin{align}\label{tehaaLL}
\left\{ \begin{gathered}
  {{\boldsymbol{\Phi }}_L} = \left[ {\begin{array}{*{20}{c}}
  {{\varphi _1}}&{{\varphi _2}}& \cdots &{{\varphi _L}} 
\end{array}} \right] = \left[ {\begin{array}{*{20}{c}}
  {{{\boldsymbol{\Phi }}_{L - 1}}}&{{\varphi _L}} 
\end{array}} \right], \hfill \\
  {{\boldsymbol{d}}_L} = {\left[ {\begin{array}{*{20}{c}}
  {{d_1} - {c_h}}&{{d_2} - {c_h}}& \cdots &{{d_h} - {c_h}} 
\end{array}} \right]^T} \hfill \\
   = {\left[ {\begin{array}{*{20}{c}}
  {{{\boldsymbol{d}}_{L - 1}}}&{{d_h} - {c_h}} 
\end{array}} \right]^T}, \hfill \\
  {{\boldsymbol{\Lambda }}_L} = \left[ {\begin{array}{*{20}{c}}
  {{{\boldsymbol{\Lambda }}_{L - 1}}}&{\boldsymbol{0}} \\ 
  {\boldsymbol{0}}&{{\theta _L}} 
\end{array}} \right], \hfill \\
  {\left[ {{{\boldsymbol{\Lambda }}_L}} \right]_{ij}} = \left\{ {\begin{array}{*{20}{l}}
  {\sum\limits_{h = 1}^H {{\lambda ^{i + h}}{H_h}{G_{\alpha ,\beta }}\left( {{e_i} - {c_h}} \right){{\left| {{e_i} - {c_h}} \right|}^{\alpha  - 2}}} ,i = j,} \\ 
  {0,i \ne j,} 
\end{array}} \right. \hfill \\
  {\theta _L} = \sum\limits_{h = 1}^H {{\lambda ^{L + h}}{H_h}{G_{\alpha ,\beta }}\left( {{e_L} - {c_h}} \right){{\left| {{e_L} - {c_h}} \right|}^{\alpha  - 2}}} . \hfill \\ 
\end{gathered}  \right.
\end{align}
To solve for the extreme values, the gradient of \eqref{respwnsmji1} is set to zero, and one can get
\begin{align}\label{dlafaiLKjia1}
{{\boldsymbol{w}}_n} = {\left( {{{\boldsymbol{\Phi }}_L}{{\boldsymbol{\Lambda }}_L}{\boldsymbol{\Phi }}_L^T + \frac{{{L^2}{\beta ^\alpha }}}{\alpha }{\vartheta _1}{\boldsymbol{I}}} \right)^{ - 1}}{{\boldsymbol{\Phi }}_L}{{\boldsymbol{\Lambda }}_L}{{\boldsymbol{d}}_L}.
\end{align}
Utilizing the kernel trick, \eqref{dlafaiLKjia1} can be rewritten as
\begin{align}
{{\boldsymbol{w}}_n} = {{\boldsymbol{\Phi }}_L}{\left( {{\boldsymbol{\Phi }}_L^T{{\boldsymbol{\Phi }}_L} + {\beta ^\alpha }{\vartheta _2}{\boldsymbol{\Lambda }}_L^{ - 1}} \right)^{ - 1}}{{\boldsymbol{d}}_L}
\end{align}
with ${\vartheta _2} = {{{L^2}{\vartheta _1}} \mathord{\left/
 {\vphantom {{{L^2}{\vartheta _1}} \alpha }} \right.
 \kern-\nulldelimiterspace} \alpha }$.

Given that the input and weight ${{\boldsymbol{w}}_n}$ are observed to combine linearly, one can obtain
\begin{align}
{{\boldsymbol{w}}_n} = {{\boldsymbol{A}}_L}{{\boldsymbol{d}}_L},
\end{align}
where ${{\boldsymbol{A}}_L} = {{\boldsymbol{\Phi }}_L}{{\boldsymbol{Q}}_L}$ and ${{\boldsymbol{Q}}_L} = {\left( {{\boldsymbol{\Phi }}_L^T{{\boldsymbol{\Phi }}_L} + {\beta ^\alpha }{\vartheta _2}{\boldsymbol{\Lambda }}_L^{ - 1}} \right)^{ - 1}}$.
Then, we can get the expression for ${\boldsymbol{Q}}_L^{ - 1}$
\begin{align}
\begin{gathered}
  {\boldsymbol{Q}}_L^{ - 1} = {\boldsymbol{\Phi }}_L^T{{\boldsymbol{\Phi }}_L} + {\beta ^\alpha }{\vartheta _2}{\boldsymbol{\Lambda }}_L^{ - 1} \hfill \\
   = \left[ {\begin{array}{*{20}{c}}
  {{\boldsymbol{\Phi }}_{L - 1}^T} \\ 
  {\varphi _L^T} 
\end{array}} \right]\left[ {\begin{array}{*{20}{c}}
  {{{\boldsymbol{\Phi }}_{L - 1}}}&{{\varphi _L}} 
\end{array}} \right] + {\beta ^\alpha }{\vartheta _2}\left[ {\begin{array}{*{20}{c}}
  {{\boldsymbol{\Lambda }}_{L - 1}^{ - 1}}&0 \\ 
  0&{\theta _L^{ - 1}} 
\end{array}} \right] \hfill \\
  {\text{ = }}\left[ {\begin{array}{*{20}{c}}
  {{\boldsymbol{Q}}_{L - 1}^{ - 1}}&{{{\boldsymbol{h}}_L}} \\ 
  {{\boldsymbol{h}}_L^T}&{\varphi _L^T{\varphi _L} + {\beta ^\alpha }{\vartheta _2}\theta _L^{ - 1}} 
\end{array}} \right], \hfill \\ 
\end{gathered} 
\end{align}
where ${{\boldsymbol{h}}_L} = {\boldsymbol{\Phi }}_{L - 1}^T{\varphi _L}$. According to the block matrix inversion, one can obtain that
\begin{align}\label{qlqljdjzlztrj1}
{{\boldsymbol{Q}}_L} = \left[ {\begin{array}{*{20}{c}}
  {{{\boldsymbol{Q}}_{L - 1}} + {{\boldsymbol{z}}_L}{\boldsymbol{z}}_L^T{\boldsymbol{r}}_L^{ - 1}}&{ - {{\boldsymbol{z}}_L}{\boldsymbol{r}}_L^{ - 1}} \\ 
  { - {\boldsymbol{z}}_L^T{\boldsymbol{r}}_L^{ - 1}}&{{\boldsymbol{r}}_L^{ - 1}} 
\end{array}} \right],
\end{align}
where ${{\boldsymbol{z}}_L} = {{\boldsymbol{Q}}_{L - 1}}{{\boldsymbol{h}}_L} = {{\boldsymbol{Q}}_{L - 1}}{\boldsymbol{\Phi }}_{L - 1}^T{\varphi _L}$ and ${{\boldsymbol{r}}_L} = \varphi _L^T{\varphi _L} + {\beta ^\alpha }{\vartheta _2}\theta _L^{ - 1} - {\boldsymbol{z}}_L^T{{\boldsymbol{h}}_L}$.
Substituting \eqref{dlafaiLKjia1} and \eqref{qlqljdjzlztrj1} into ${{\boldsymbol{A}}_L} = {{\boldsymbol{\Phi }}_L}{{\boldsymbol{Q}}_L}$, and we can get
\begin{align}\label{Aldljchrjlaljd}
\begin{gathered}
  {{\boldsymbol{A}}_L} = \left[ {\begin{array}{*{20}{c}}
  {{{\boldsymbol{Q}}_{L - 1}} + {{\boldsymbol{z}}_L}{\boldsymbol{z}}_L^T{\boldsymbol{r}}_L^{ - 1}}&{ - {{\boldsymbol{z}}_L}{\boldsymbol{r}}_L^{ - 1}} \\ 
  { - {\boldsymbol{z}}_L^T{\boldsymbol{r}}_L^{ - 1}}&{{\boldsymbol{r}}_L^{ - 1}} 
\end{array}} \right]{{\boldsymbol{d}}_L} \hfill \\
   = \left[ {\begin{array}{*{20}{c}}
  {{{\boldsymbol{Q}}_{L - 1}} + {{\boldsymbol{z}}_L}{\boldsymbol{z}}_L^T{\boldsymbol{r}}_L^{ - 1}}&{ - {{\boldsymbol{z}}_L}{\boldsymbol{r}}_L^{ - 1}} \\ 
  { - {\boldsymbol{z}}_L^T{\boldsymbol{r}}_L^{ - 1}}&{{\boldsymbol{r}}_L^{ - 1}} 
\end{array}} \right]\left[ {\begin{array}{*{20}{c}}
  {{{\boldsymbol{d}}_{L - 1}}} \\ 
  {{d_L} - {c_h}} 
\end{array}} \right] \hfill \\
   = \left[ {\begin{array}{*{20}{c}}
  {{{\boldsymbol{A}}_{L - 1}} - {{\boldsymbol{z}}_L}{\boldsymbol{r}}_L^{ - 1}\left( {{d_L} - {c_h}} \right)} \\ 
  {{\boldsymbol{r}}_L^{ - 1}\left( {{d_L} - {c_h}} \right)} 
\end{array}} \right]. \hfill \\ 
\end{gathered} 
\end{align}

According to the above-detailed derivation, the pseudo-code of the proposed QKGMEE algorithm is summarised in Algorithm \ref{algorithmqkrgmee}.
\begin{algorithm}[t]
\caption{QKRGMEE} \label{algorithmqkrgmee}
\LinesNumbered 
\KwIn{sample sequences ${\left\{ {{d_n},{{\boldsymbol{u}}_n}} \right\},n = 1,2, \cdots }$}
\KwOut{function ${\boldsymbol{f}}( \cdot )$}
\textbf{Parameters setting}: select the proper parameters including $\gamma$, ${\alpha}$, and ${\beta}$\; 
\textbf{Initialization}: ${{\boldsymbol{Q}}_1} = {\left[ {{\beta ^\alpha }{\vartheta _2} + \kappa \left( {{{\boldsymbol{u}}_1},{{\boldsymbol{u}}_1}} \right)} \right]^{ - 1}}$, ${{\boldsymbol{A}}_1} = {{\boldsymbol{Q}}_1}{d_1}$\; 
\While{${\left\{ {{d_n},{{\boldsymbol{u}}_n}} \right\} \ne \emptyset }$}
{
Compute ${{{\boldsymbol{h}}_L}}$, ${{y_L}}$, and ${{e_L}}$ by
\begin{subequations}
\begin{numcases}{}
\begin{gathered}
  {{\boldsymbol{h}}_L} =  \hfill \\
  {\left[ {\kappa \left( {{u_{n + L - 1}},{u_n}} \right), \cdots \kappa \left( {{u_{n + L - 1}},{u_{n + L - 2}}} \right)} \right]^T}, \hfill \\ 
\end{gathered} \\
{y_L} = {\boldsymbol{h}}_L^{\text{T}}{{\boldsymbol{A}}_{L - 1}},\\
{e_L} = {d_L} - {y_L};
\end{numcases}
\end{subequations}\

Compute ${{{\boldsymbol{z}}_L}}$, ${{\theta _L}}$, ${{{\boldsymbol{r}}_L}}$\;
Update ${{{\boldsymbol{Q}}_L}}$ and ${{{\boldsymbol{A}}_L}}$\;
}
\end{algorithm}

\begin{algorithm}[t]
\caption{QKRMEE} \label{algorithmqkrmee}
\LinesNumbered 
\KwIn{sample sequences ${\left\{ {{d_n},{{\boldsymbol{u}}_n}} \right\},n = 1,2, \cdots }$}
\KwOut{function ${\boldsymbol{f}}( \cdot )$}
\textbf{Parameters setting}: select the proper parameters $\gamma$ and ${\sigma}$\; 
\textbf{Initialization}: ${{\boldsymbol{Q}}_{1:S}} = {\left[ {{\gamma ^2}{\vartheta _{2;S}} + \kappa \left( {{{\boldsymbol{u}}_1},{{\boldsymbol{u}}_1}} \right)} \right]^{ - 1}}$, (where ${\vartheta _{2;S}} = {L^2}{\vartheta _1}$), ${{\boldsymbol{A}}_{1;S}} = {{\boldsymbol{Q}}_{1;S}}{d_1}$\; 
\While{${\left\{ {{d_n},{{\boldsymbol{u}}_n}} \right\} \ne \emptyset }$}
{
Compute ${{{\boldsymbol{h}}_L}}$, ${y_{L;S}}$, and ${e_{L;S}}$ by
\begin{subequations}
\begin{numcases}{}
\begin{gathered}
  {{\boldsymbol{h}}_L} =  \hfill \\
  {\left[ {\kappa \left( {{u_{n + L - 1}},{u_n}} \right), \cdots \kappa \left( {{u_{n + L - 1}},{u_{n + L - 2}}} \right)} \right]^T}, \hfill \\ 
\end{gathered} \\
{y_{L;S}} = {\boldsymbol{h}}_L^T{{\boldsymbol{A}}_{L - 1;S}},\\
{e_{L;S}} = {d_L} - {y_{L;S}};
\end{numcases}
\end{subequations}\

Compute ${{\boldsymbol{z}}_{L;S}}$, ${\theta _{L;S}}$, ${{\boldsymbol{r}}_{L;S}}$ using
\begin{align}\label{rlathetals}
\left\{ \begin{gathered}
  {{\boldsymbol{z}}_{L;S}} = {{\boldsymbol{Q}}_{L - 1;S}}{{\boldsymbol{h}}_L} = {{\boldsymbol{Q}}_{L - 1;S}}{\boldsymbol{\Phi }}_{L - 1}^T{\varphi _L}, \hfill \\
  {\theta _{L;S}} = \sum\limits_{h = 1}^H {{\lambda ^{L + h}}{H_h}{G_\sigma }\left( {{e_L} - {c_h}} \right)} , \hfill \\
  {{\boldsymbol{r}}_{L;S}} = \varphi _L^T{\varphi _L} + {\sigma ^2}{\vartheta _{2;S}}\theta _{L;S}^{ - 1} - {\boldsymbol{z}}_{L;S}^T{{\boldsymbol{h}}_L}; \hfill \\ 
\end{gathered}  \right.
\end{align}
Update ${{\boldsymbol{Q}}_{L;S}}$ and ${{\boldsymbol{A}}_{L;S}}$ using 
\begin{align}
{{\boldsymbol{Q}}_{L;S}} = \left[ {\begin{array}{*{20}{c}}
  {{{\boldsymbol{Q}}_{L - 1;S}} + {{\boldsymbol{z}}_{L;S}}{\boldsymbol{z}}_{L;S}^T{\boldsymbol{r}}_{L;S}^{ - 1}}&{ - {{\boldsymbol{z}}_{L;S}}{\boldsymbol{r}}_{L;S}^{ - 1}} \\ 
  { - {\boldsymbol{z}}_{L;S}^T{\boldsymbol{r}}_{L;S}^{ - 1}}&{{\boldsymbol{r}}_{L;S}^{ - 1}} 
\end{array}} \right],
\end{align}
and
\begin{align}
{{\boldsymbol{A}}_{L;S}} = \left[ {\begin{array}{*{20}{c}}
  {{{\boldsymbol{A}}_{L - 1;S}} - {{\boldsymbol{z}}_{L;S}}{\boldsymbol{r}}_{L;S}^{ - 1}\left( {{d_L} - {c_h}} \right)} \\ 
  {{\boldsymbol{r}}_{L;S}^{ - 1}\left( {{d_L} - {c_h}} \right)} 
\end{array}} \right];{\text{ }}
\end{align}
}
\end{algorithm}

\begin{remark}
When $\gamma=0$, the proposed QKRGMEE algorithm translates into KRGMEE algorithm \cite{HE2023109188}. It can be observed that the KRGMEE algorithm is a special form of the QKRGMEE algorithm, and the QKRGMEE algorithm has a much smaller computational burden.
\end{remark}

It is obvious to infer that the QKRGMEE algorithm will translate into a special algorithm with a QMEE cost function, and the derived algorithm is known to us as QKRMEE. The QKRMEE algorithm and the QKRGMEE algorithm share a similar comprehensive derivation procedure. The QKRMEE derivation method is skipped to cut down on repetition, while Algorithm \ref{algorithmqkrmee} provides an overview of its pseudo-code.

\begin{remark}
The QKRGMEE algorithm translates into the proposed QKRMEE algorithm for $\alpha=2$. When $\alpha=2$ and $\gamma=0$, the proposed QKRGMEE algorithm translates into the KRMEE algorithm.
\end{remark}

\section{Performance analysis}\label{perfroanaly}
\subsection{Mean error behavior}
By using the QGMEE criterion, the output of the nonlinear system's desired result is
\begin{align}\label{Vnjfainwot}
{d_n} = {\left( {{{\boldsymbol{w}}^o}} \right)^T}{{\boldsymbol{\varphi }}_n} + {v_n},
\end{align}
where ${{\boldsymbol{w}}^o}$ represents the unknown parameter and ${v_n}$ denotes the zero-mean measurement noise. The weight ${\boldsymbol{w}}$ can also, from \cite{wang2021quaternion}, be expressed as
\begin{align}\label{enrnjianaznj}
{{\boldsymbol{w}}_n} = {{\boldsymbol{w}}_{n - 1}} + \left( {{{\boldsymbol{\varphi }}_n} - {{\boldsymbol{\Phi }}_{n - 1}}{{\boldsymbol{z}}_n}} \right){\boldsymbol{r}}_n^{ - 1}{{\boldsymbol{e}}_n}.
\end{align}

Suppose that the weight error definition is
\begin{align}\label{errorsimnfenhao}
{{\boldsymbol{\varepsilon }}_n} = {{\boldsymbol{w}}^o} - {{\boldsymbol{w}}_n},
\end{align}
where ${{\boldsymbol{\varepsilon }}_n}$ is a vector ${{\boldsymbol{\varepsilon }}_n} = {\left[ {\begin{array}{*{20}{c}}
  {{\varepsilon _{1;n}}}&{{\varepsilon _{2;n}}}& \cdots &{{\varepsilon _{m;n}}} 
\end{array}} \right]^T}$. Substituting \eqref{enrnjianaznj} into \eqref{errorsimnfenhao}, and we can obtain
\begin{align}\label{enjuhaornjain1}
{{\boldsymbol{\varepsilon }}_n} = {{\boldsymbol{w}}^o} - {{\boldsymbol{w}}_n} = {{\boldsymbol{\varepsilon }}_{n - 1}} - \left( {{{\boldsymbol{\varphi }}_n} - {{\boldsymbol{\Phi }}_{n - 1}}{{\boldsymbol{z}}_n}} \right){\boldsymbol{r}}_n^{ - 1}{{\boldsymbol{e}}_n}.
\end{align}
According to \eqref{Vnjfainwot}, one can obtain 
\begin{align}\label{efwowvnjft}
{{\boldsymbol{e}}_n} = {\boldsymbol{\varphi }}_n^T{{\boldsymbol{w}}^o} + {v_n} - {\boldsymbol{\varphi }}_n^T{{\boldsymbol{w}}_{n - 1}}.
\end{align}
Substituting \eqref{efwowvnjft} into \eqref{enjuhaornjain1}, and we can get
\begin{align}
{{\boldsymbol{\varepsilon }}_n} = \left( {{\boldsymbol{I}} - {{\boldsymbol{\alpha }}_n}{\boldsymbol{\varphi }}_n^T} \right){{\boldsymbol{\varepsilon }}_{n - 1}} - {{\boldsymbol{\alpha }}_n}{v_n},
\end{align}
where ${{\boldsymbol{\alpha }}_n} = \left( {{{\boldsymbol{\varphi }}_n} - {{\boldsymbol{\Phi }}_{n - 1}}{{\boldsymbol{z}}_n}} \right){\boldsymbol{r}}_n^{ - 1}$. Given that ${v_n}$'s mean value is zero, the expectation of ${{\boldsymbol{\varepsilon }}_n}$ can be written as
\begin{align}\label{36nedssiger}
E\left[ {{{\boldsymbol{\varepsilon }}_n}} \right] = \left( {{\boldsymbol{I}} - E\left[ {{{\boldsymbol{\alpha }}_n}{\boldsymbol{\varphi }}_n^T} \right]} \right)E\left[ {{{\boldsymbol{\varepsilon }}_{n - 1}}} \right].
\end{align}
The eigenvalue decomposition of ${E\left[ {{{\boldsymbol{\alpha }}_n}{\boldsymbol{\varphi }}_n^T} \right]}$ is $E\left[ {{{\boldsymbol{\alpha }}_n}{\boldsymbol{\varphi }}_n^T} \right] = {\boldsymbol{{\rm K}\Omega }}{{\boldsymbol{{\rm K}}}^T}$. ${\boldsymbol{{\rm K}}}$ denotes a square matrix composed of eigenvectors whose diagonal elements are eigenvalues. When we set ${\boldsymbol{\varepsilon }} = {{\boldsymbol{{\rm K}}}^T}{\boldsymbol{\varepsilon }}$, \eqref{36nedssiger} can be further written as
\begin{align}
E\left[ {{{\boldsymbol{\varepsilon }}_n}} \right] = \left( {{\boldsymbol{I}} - {\boldsymbol{\Omega }}} \right)E\left[ {{{\boldsymbol{\varepsilon }}_{n - 1}}} \right].
\end{align}
As a result, W's maximum eigenvalue $E\left[ {{{\boldsymbol{\alpha }}_n}{\boldsymbol{\varphi }}_n^T} \right]$ is less than one, which implies that $E\left[ {{{\boldsymbol{\varepsilon }}_n}} \right]$ will eventually converge.
\subsection{Mean square error behavior}
As the noise ${v_n}$ is never correlated with the ${{{\boldsymbol{\varepsilon }}_{n - 1}}}$, the covariance matrix of
\begin{align}\label{EnntIjinasafaer}
\begin{gathered}
  E\left[ {{{\boldsymbol{\varepsilon }}_n}{\boldsymbol{\varepsilon }}_n^T} \right] = {{\boldsymbol{\alpha }}_n}E\left[ {{v_n}{v_n}} \right]{\boldsymbol{\alpha }}_n^T{\text{ + }} \hfill \\
  \left( {{\boldsymbol{I}} - {{\boldsymbol{\alpha }}_n}{\boldsymbol{\varphi }}_n^T} \right)E\left[ {{{\boldsymbol{\varepsilon }}_{n - 1}}{\boldsymbol{\varepsilon }}_{n - 1}^T} \right]{\left( {{\boldsymbol{I}} - {{\boldsymbol{\alpha }}_n}{\boldsymbol{\varphi }}_n^T} \right)^T}. \hfill \\ 
\end{gathered} 
\end{align}
\eqref{EnntIjinasafaer} can be abbreviated as
\begin{align}\label{trtretySheng}
{{\boldsymbol{T}}_n} = {{\boldsymbol{R}}_n}{{\boldsymbol{T}}_{n - 1}}{\boldsymbol{R}}_n^T + {{\boldsymbol{\Xi }}_n}
\end{align}
with
\begin{align}
\left\{ \begin{gathered}
  {{\boldsymbol{T}}_n} = E\left[ {{{\boldsymbol{\varepsilon }}_n}{\boldsymbol{\varepsilon }}_n^T} \right], \hfill \\
  {{\boldsymbol{R}}_n} = \left( {{\boldsymbol{I}} - {{\boldsymbol{\alpha }}_n}{\boldsymbol{\varphi }}_n^T} \right), \hfill \\
  {{\boldsymbol{\Xi }}_n} = {{\boldsymbol{\alpha }}_n}E\left[ {{v_n}{v_n}} \right]{\boldsymbol{\alpha }}_n^T. \hfill \\ 
\end{gathered}  \right.
\end{align}
Given that, ${{{\boldsymbol{\alpha }}_n}{\boldsymbol{\varphi }}_n^T}$ and ${{\boldsymbol{\alpha }}_n}$ are variables that are independent of time \cite{WANG2022108410}, one can summarize
\begin{align}
\left\{ \begin{gathered}
  \mathop {\lim }\limits_{n \to \infty } {{\boldsymbol{T}}_n} = {\boldsymbol{T}}, \hfill \\
  \mathop {\lim }\limits_{n \to \infty } {{\boldsymbol{R}}_n} = {\boldsymbol{R}}, \hfill \\
  \mathop {\lim }\limits_{n \to \infty } {{\boldsymbol{\Xi }}_n} = {\boldsymbol{\Xi }}. \hfill \\ 
\end{gathered}  \right.
\end{align}
As a result, when $n \to \infty $, \eqref{trtretySheng} can be written as a real discrete-time Lyapunov equation with the following formula:
\begin{align}\label{trtrTsanheng}
{\boldsymbol{T}} = {\boldsymbol{RT}}{{\boldsymbol{R}}^T} + {\boldsymbol{\Xi }}.
\end{align}
From the matrix-vector operator:
\begin{align}
\left\{ \begin{gathered}
  {\text{vec}}\left( {{\boldsymbol{SUV}}} \right) = \left( {{{\boldsymbol{V}}^{\text{T}}} \otimes {\boldsymbol{S}}} \right){\text{vec}}\left( {\boldsymbol{U}} \right), \hfill \\
  {\text{vec}}\left( {{\boldsymbol{S}} + {\boldsymbol{V}}} \right) = {\text{vec}}\left( {\boldsymbol{S}} \right) + {\text{vec}}\left( {\boldsymbol{V}} \right), \hfill \\ 
\end{gathered}  \right.
\end{align}
where $ \otimes $ represents the Kronecker product and ${\text{vec}}\left(  \cdot  \right)$ stands for vectorization operation.
The closed-form solution of \eqref{trtrTsanheng} can be written as
\begin{align}
{\text{vec}}\left( {\boldsymbol{T}} \right) = {\left( {{\boldsymbol{I}} - {\boldsymbol{R}} \otimes {\boldsymbol{R}}} \right)^{ - 1}}vec\left( {\boldsymbol{\Xi }} \right).
\end{align}

\subsection{Computational Complexity}
In comparison to the KRMEE and KRGMEE algorithms, the computational complexity of the proposed QKRMEE and QKRGMEE algorithms is examined. By comparing the pseudocode of these algorithms, it can be deduced that the formulas involved in these algorithms have the same form except for the calculation of ${\theta _{L;S}}$ and ${\theta _L}$. The method in \cite{CHEN201770} for assessing the computational burden of each algorithm is used to make it easier to compare the computational complexity of different algorithms. The difference between the KRMEE and QKRMEE algorithms' computational complexity can therefore be stated as
\begin{align}
\left\{ \begin{gathered}
  {C_{QKRMEE}} = {C_{com}} + {C_{\theta ;QKRMEE}}, \hfill \\
  {C_{KRMEE}} = {C_{com}} + {C_{\theta ;KRMEE}}, \hfill \\ 
\end{gathered}  \right.
\end{align}
where ${C_{KRMEE}}$ and ${C_{QKRMEE}}$ are the computational complexity of one cycle of the KRMEE and QKRMEE algorithms; ${C_{com}}$ represents the computational complexity of formulas of the same form in both algorithms; ${C_{\theta ;KRMEE}}$ and ${C_{\theta ;QKRMEE}}$ are the computational complexity of ${\phi _L}$ in (17) \cite{WANG2022108410} and ${\theta _{L;S}}$ in \eqref{rlathetals}. The difference ${C_{d;MEE}}$ in computational complexity between the KRMEE and QKRMEE algorithms can be expressed as
\begin{align}
{C_{d;MEE}} = {C_{\theta ;KRMEE}} - {C_{\theta ;QKRMEE}}.
\end{align}
Similarly, the difference ${C_{d;GMEE}}$ in computational complexity between the KRGMEE and QKRGMEE algorithms can be expressed as
\begin{align}
{C_{d;GMEE}} = {C_{\theta ;KRGMEE}} - {C_{\theta ;QKRGMEE}},
\end{align}
where ${C_{\theta ;KRGMEE}}$ and ${C_{\theta ;QKRGMEE}}$ are the computational complexity of ${\psi _L}$ in (30g) \cite{HE2023109188} and ${\theta _{L}}$ in \eqref{tehaaLL}. The computational complexity of ${C_{\theta ;KRMEE}}$, ${C_{\theta ;QKRMEE}}$, ${C_{\theta ;KRGMEE}}$, and ${C_{\theta ;QKRGMEE}}$ is shown in Table \ref{COMPULLSFAI}. Based on the estimation scheme of the computational complexity in \cite{CHEN201770}, one can obtain that
\begin{align}\label{4820H16Hj14}
\left\{ \begin{gathered}
  {C_{d;MEE}} \approx 15L - 14 - 16H, \hfill \\
  {C_{d;GMEE}} \approx 19L - 18 - 20H. \hfill \\ 
\end{gathered}  \right.
\end{align}

\begin{remark}
The reduced computational burden after quantization by approximating the strategy in \cite{CHEN201770}, \eqref{4820H16Hj14} can reflect the contribution of the quantification mechanism to a certain extent, it is not completely accurate. From \eqref{4820H16Hj14}, the quantization approach can successfully lessen the computational burden on the KRMEE and KRGMEE algorithms when $L$ is big and $L \ll H$. It is worth noting that reducing the computational burden will, to some extent, decrease the steady-state error performance of the suggested algorithms. How to choose the quantization threshold to trade off the performance of the algorithm against the computational complexity is discussed in detail in Section \ref{simulation}.
\end{remark}

\begin{table}
\centering
\caption{The computational complexity of key components.}\label{COMPULLSFAI}
\begin{tabular}{llll}
\hline
  & $ \times / \div $ & $ + / - $    & Exponentiation      \\ \hline
${C_{\theta ;KRMEE}}$ & $8L - 8$   & $3L - 3$ & $4L - 4$ \\
${C_{\theta ;QKRMEE}}$ & $9H$       & $3H-1$   & $4H$ \\
${C_{\theta ;KRGMEE}}$ & $9L - 9$   & $4L-4$   & $6L - 6$ \\
${C_{\theta ;QKRGMEE}}$ & $10H$      & $4L-1$   & $6H$ \\ \hline
\end{tabular}
\end{table}
\section{Simulations}\label{simulation}
To demonstrate the effectiveness of the QKRGMEE and QKRMEE algorithms, we present various simulations, and the MSE is regarded as a tool to measure the algorithm performance in terms of steady-state error. Several noise models covered in this paper are presented before these simulations are implemented, such as mixed-Gaussian noise, Gaussian noise, Rayleigh noise, etc.
\begin{enumerate} 
\item The mixed-Gaussian model \cite{9923771} takes the following form:
\begin{align}
v \sim \varsigma \mathcal{N}\left( {{a_1},{\mu _1}} \right) + \left( {1 - \varsigma } \right)\mathcal{N}\left( {{a_2},{\mu _2}} \right),0 \leqslant \varsigma  \leqslant 1,
\end{align}
where $\mathcal{N}\left( {{a_1},{\mu _1}} \right){\text{ }}$ denotes the Gaussian distribution with mean $a_1$ and variance ${{\mu _1}}$, and $\varsigma  $ represents the mixture coefficient of two kinds of Gaussian distribution. The mixed-Gaussian distribution can be abbreviated as $v \sim M\left( {\varsigma ,{a_1},{a_2},{\mu _1},{\mu _2}} \right)$.
\item The Rayleigh distribution's probability density function is written as $r\left( t \right) = \left( {{t \mathord{\left/
 {\vphantom {t {{\chi ^2}}}} \right.
 \kern-\nulldelimiterspace} {{\chi ^2}}}} \right)\exp \left( {{{ - {t^2}} \mathord{\left/
 {\vphantom {{ - {t^2}} {2{\chi ^2}}}} \right.
 \kern-\nulldelimiterspace} {2{\chi ^2}}}} \right)$. The noise that follows a Rayleigh distribution is shown as $v \sim R\left( \chi  \right)$.
\end{enumerate}

In this paper, four scenarios are considered, and the distribution of the noise for these four scenarios is $R\left( 3 \right)$, $M\left( {0.95,0,0,0.01,64} \right)$, $\mathcal{N}\left( {0,0.01} \right)$, and $0.2R\left( 3 \right) + 0.8M\left( {0.8,0,0,0.01,64} \right)$, respectively.

\subsection{Mackey–Glass time series prediction}
This subpart tests the QKRMEE and QKRGMEE algorithms' performance to learn nonlinearly using the benchmark data set known as Mackey-Glass (MG) chaotic time series. A nonlinear delay differential equation known as the MG equation has the following form:
\begin{align}
\frac{{ds\left( t \right)}}{{dt}} = \frac{{0.2s\left( {t - \tau } \right)}}{{1 + {s^{10}}\left( {t - \tau } \right)}} - 0.1s\left( t \right).
\end{align}
By resolving the MG equation, 1000 noise-added training data and 100 test data are produced. 

\begin{figure*}[htbp] 
\centering  
\subfigure[Rayleigh noises.]{
\includegraphics[width=0.46\textwidth]{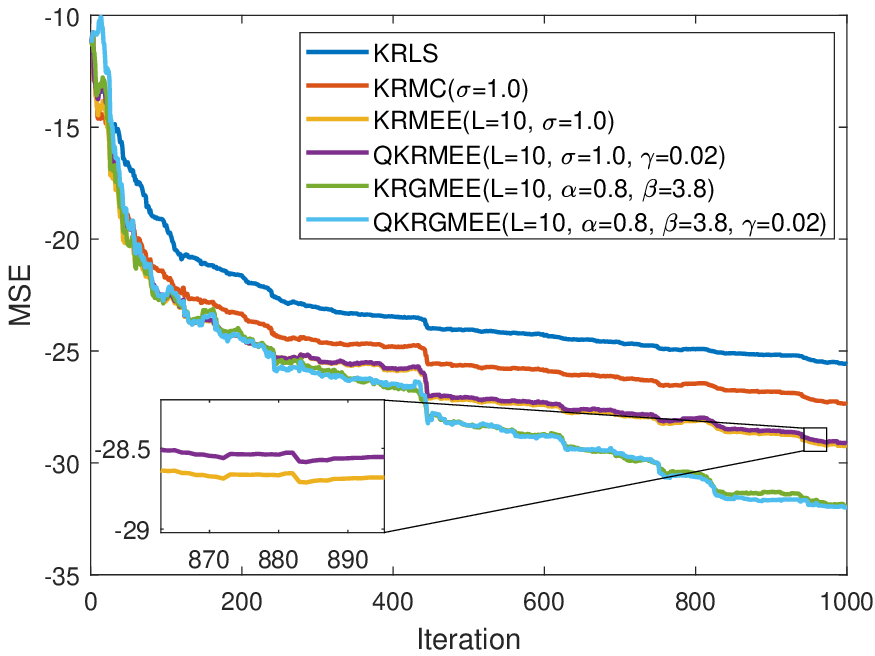}\label{qKRMEE}
}
\quad
\subfigure[Mixed-Gaussian noise.]{
\includegraphics[width=0.46\textwidth]{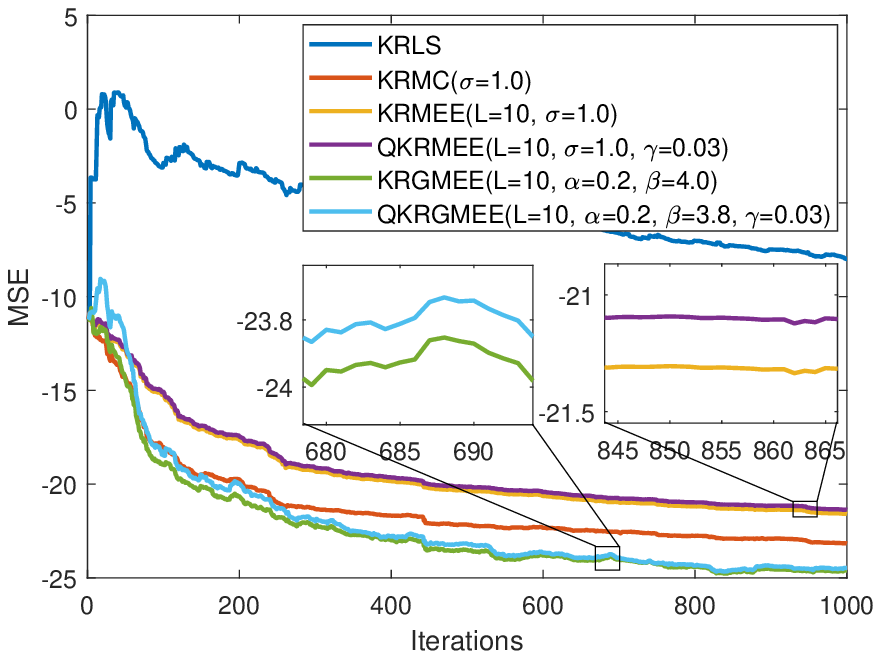}\label{mixed_Gaussian_qkrgmee}
}
\quad
\subfigure[Gaussian noise.]{
\includegraphics[width=0.46\textwidth]{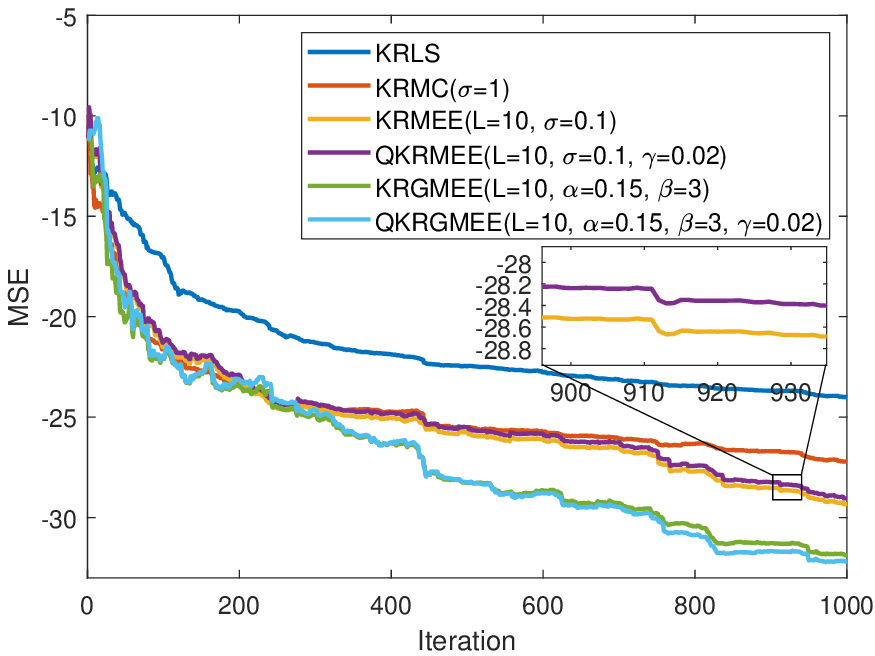}\label{fig_Gaussian}
}
\quad
\subfigure[Mixed-noise (fourth scenario).]{
\includegraphics[width=0.46\textwidth]{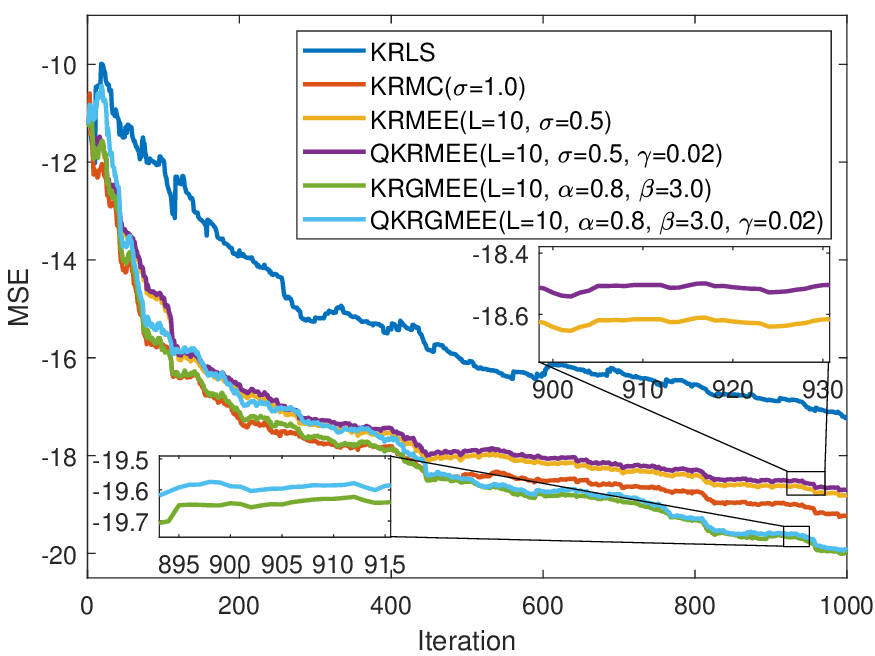}\label{fig_Laplace}
}
\caption{Convergence curves under different scenarios}\label{convercurnoise}
\end{figure*}

In the four aforementioned instances, the performance of the QKRGMEE and QKRMEE algorithms is compared with that of the KRLS \cite{engel2004kernel}, KRMC \cite{WU201511}, KRMEE \cite{WANG2022108410}, and KRGMEE \cite{HE2023109188} algorithms. In Fig. \ref{convercurnoise}, the parameters of the algorithms and the MSE convergence curves are displayed, and the regularization factors of KRLS-type adaptive filtering algorithms are all 1. It is evident that the proposed QKRMEE and QKRGMEE algorithms perform marginally worse than the KRMEE and KRGMEE algorithms. 

\subsection{The relationship between parameters and performance}
In this section, we investigated how the shape parameter $\alpha $, scale parameter $\beta $, length of the Parzen window $L$, and quantization threshold $\gamma$ of the QKRGMEE algorithm affected performance on the performance in terms of MSE. Since the QKRMEE algorithm is a special form of the QKRGMEE algorithm, one focuses on the influence of parameters on the performance of the QKRGMEE algorithm. The discussion of how the parameter settings affect the functionality of the QKRGMEE algorithm continues to use the MG chaotic time series. The results reached can also serve as a guide for choosing the QKRGMEE algorithm's parameters.

\begin{figure*}[htbp] 
\centering  
\subfigure[Parameter L.]{
\includegraphics[width=0.46\textwidth]{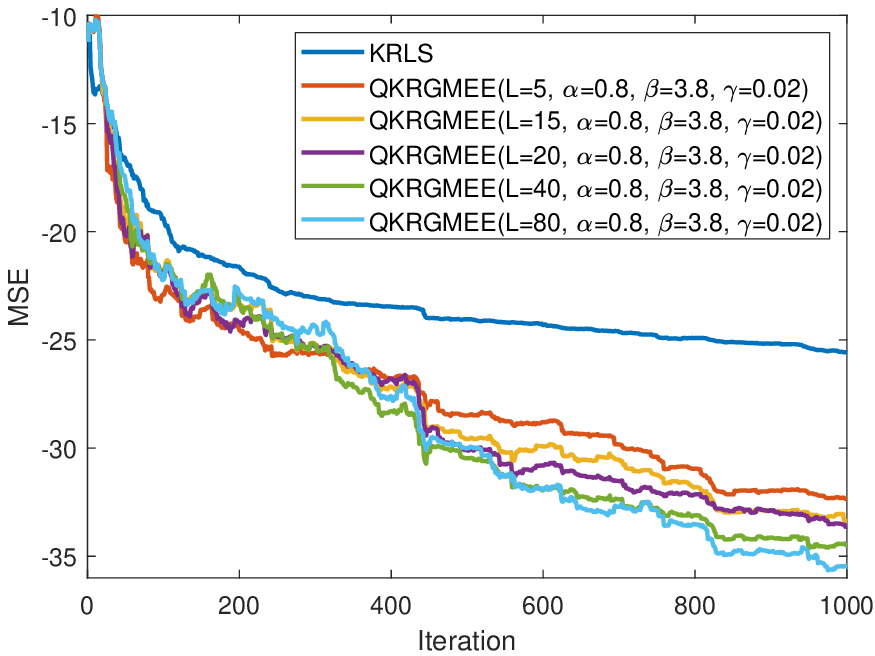}\label{qKRMEE_para_L}
}
\quad
\subfigure[Parameter $\gamma$.]{
\includegraphics[width=0.46\textwidth]{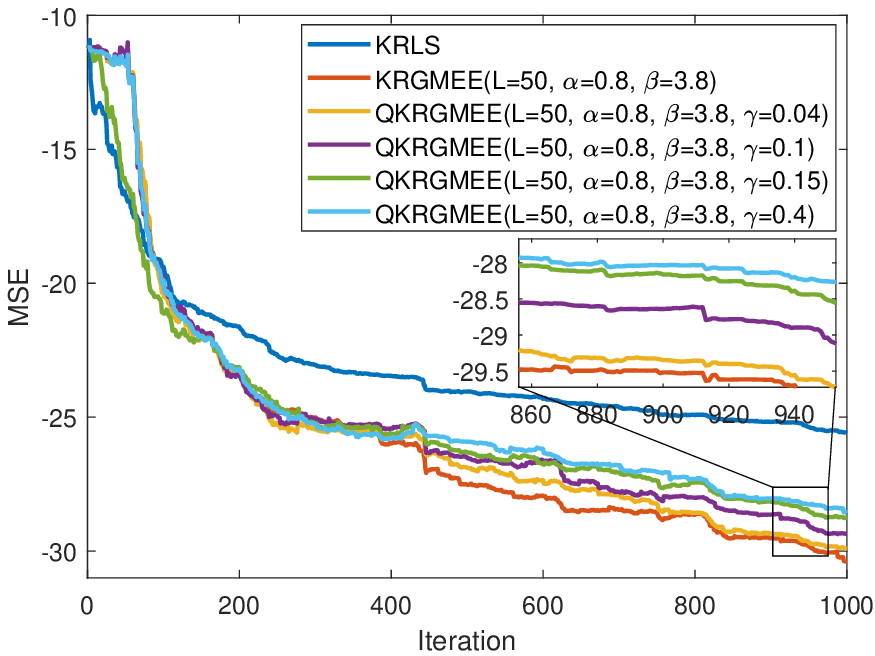}\label{qKRMEE_para_gamma}
}
\quad
\subfigure[Parameter $\alpha$.]{
\includegraphics[width=0.46\textwidth]{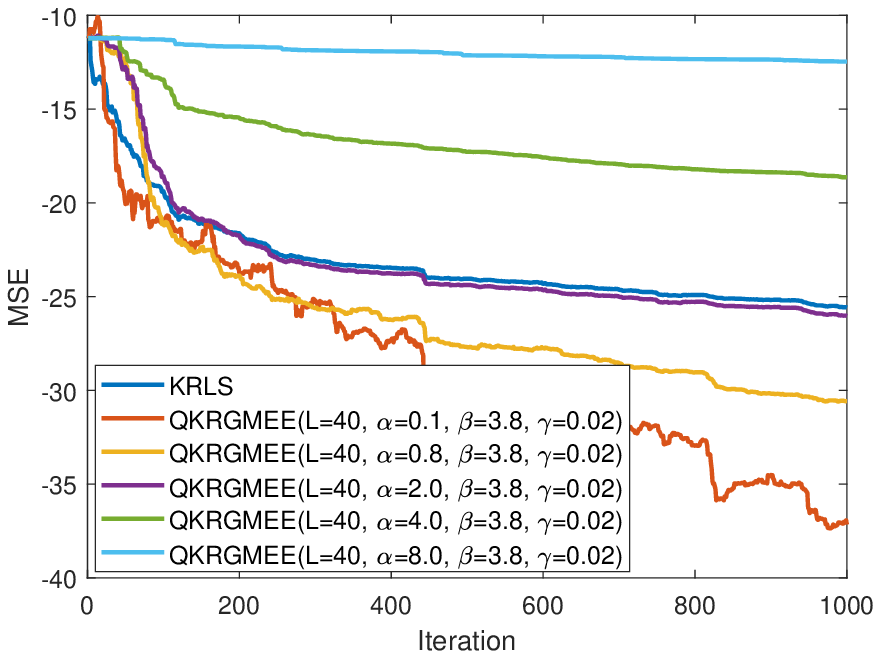}\label{qKRMEE_para_alpha}
}
\quad
\subfigure[Parameter $\beta$.]{
\includegraphics[width=0.46\textwidth]{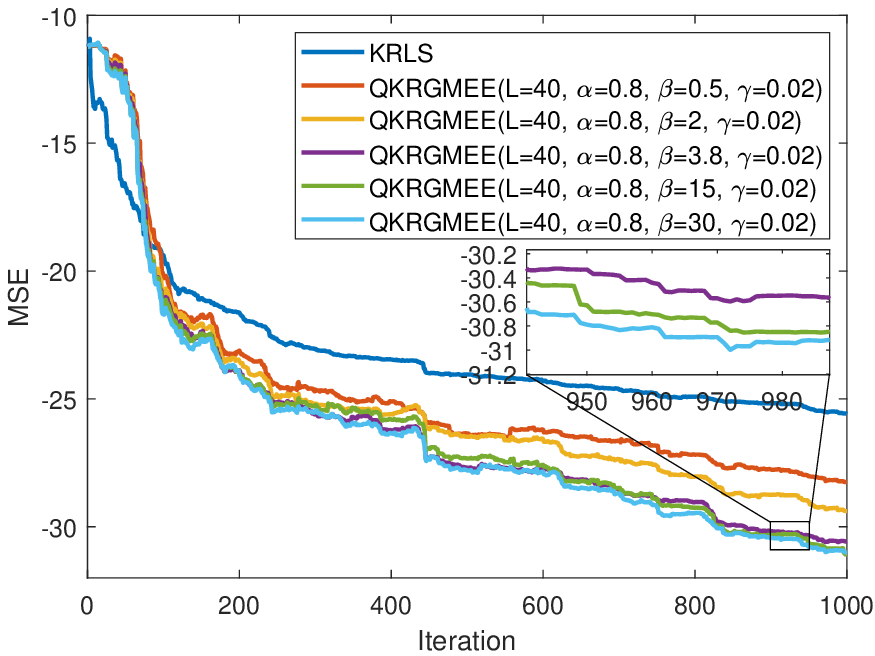}\label{qKRMEE_para_beta}
}
\caption{The performance of the proposed QKRGMEE algorithm with different parameters}\label{parameter}
\end{figure*}

First, the values of these parameters are shown in Fig. \ref{qKRMEE_para_L} as we investigate the impact of parameter ${L}$ on the functionality of the QKRGMEE algorithm. The parameter ${L}$ is set to $L = 5,15,20,40,80$ in this simulation. The simulation results are shown in Fig. \ref{qKRMEE_para_L} and Table \ref{diff_noise_L}, and the distribution of the additive noise is the same as it was in the prior simulation. Fig. \ref{qKRMEE_para_L} shows the convergence curves of the QKRGMEE algorithm with different $L$ in the first scenario. Table \ref{diff_noise_L} presents the steady-state MSE with different $L$ and scenarios. Simulations show that the proposed QKRGMEE algorithms' steady-state error lowers as $L$ increases with the four noise categories listed above. In addition, the improvement in terms of performance is not significant when $L$ is greater than 50, thus, it is possible to balance the performance of the algorithm with the amount of computation when $L$ is less than 50. 

Second, the influence of the quantization threshold on the performance of the algorithm is shown in Fig. \ref{qKRMEE_para_gamma} and Table \ref{diff_noise_gamma_time}. Fig. \ref{qKRMEE_para_gamma} presents the convergence curve of the MSE with different $\gamma$ with the presence of Rayleigh noise, and $L$ is set as $L=50$. Table \ref{diff_noise_gamma_time} shows the MSE, the running time of each iteration, and the number of elements $H$ in the quantized error set with the different $\gamma$ and scenarios. These KRLS algorithms are measured using MATLAB 2020a, which works on an i5-8400 and a 2.80GHz CPU. Moreover, the KRLS and KRGMEE algorithms are used as benchmarks. From these simulation results, it can be inferred that both the running time and the number $H$ decrease as the quantization threshold increases, while the performance of the algorithm also decreases to some extent; moreover, one can obtain $H \ll L$. The suggested range of quantization thresholds is $0.04 \leqslant \gamma  \leqslant 0.15$, which strikes a balance between the QKRGMEE algorithm's efficiency and computing complexity.

\begin{figure*}[htbp] 
\centering  
\subfigure[Rayleigh noise]{
\includegraphics[width=0.46\textwidth]{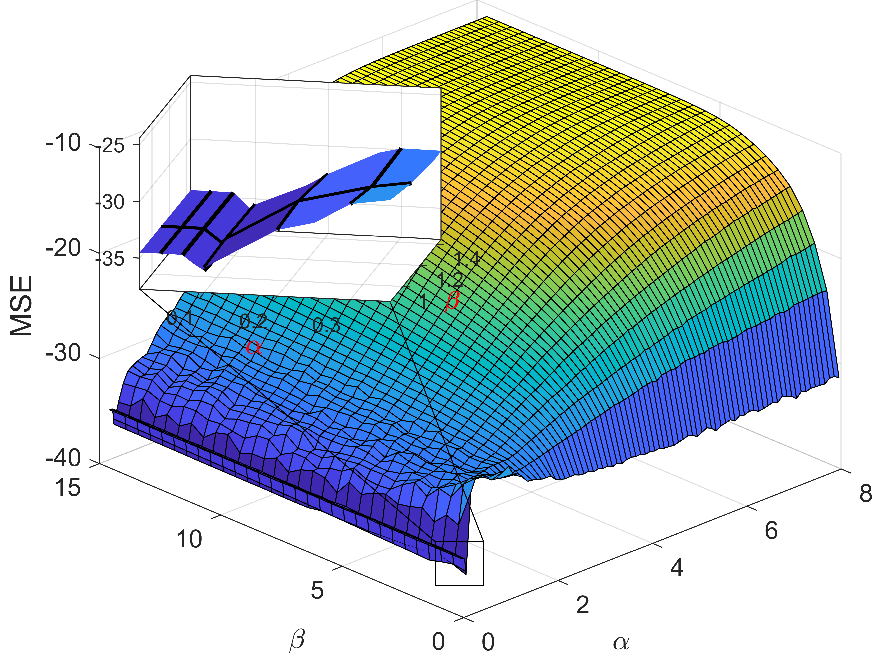}\label{KRGMEE_MSE_RAY_AB}
}
\quad
\subfigure[Mixed-Gaussian noise]{
\includegraphics[width=0.46\textwidth]{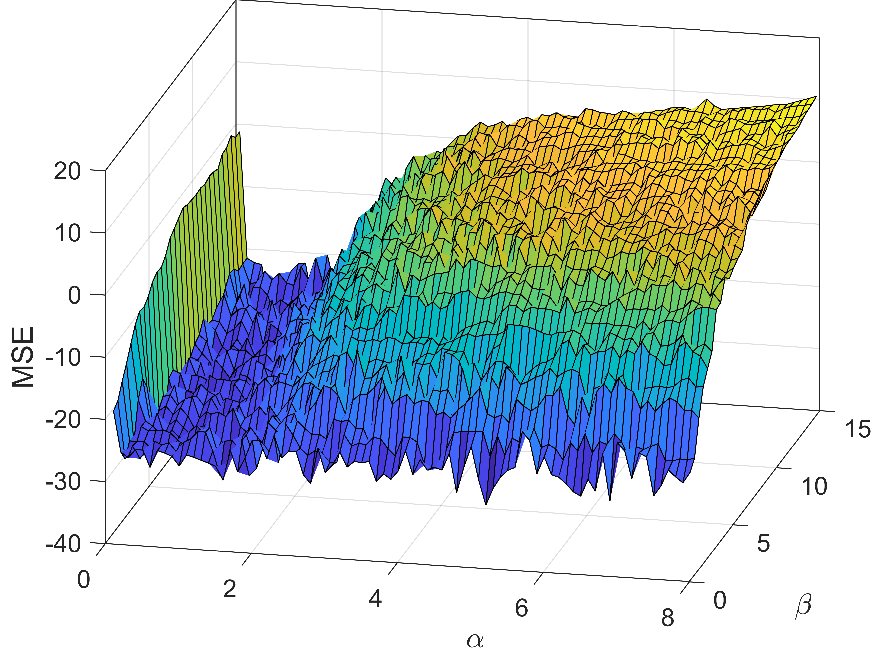}\label{mixedagaussianalbe}
}
\caption{The performance surfaces of the QKRGMEE algorithm with respect to the parameters $\alpha$ and $\beta$ in different scenarios}\label{perfor_alpha_Ray_mixed_noise}
\end{figure*}

Final, it is also addressed how the parameters $\alpha$ and $\beta$ affect the QKRGMEE algorithm's performance. The simulation results are presented in Fig. \ref{qKRMEE_para_alpha}, Fig. \ref{qKRMEE_para_beta}, Fig. \ref{perfor_alpha_Ray_mixed_noise}, Table \ref{diff_noise_alpha_time}, and Table \ref{diff_noise_beta_time}. Fig. \ref{qKRMEE_para_alpha} and Fig. \ref{qKRMEE_para_beta} show, respectively, the convergence curves of the steady-state MSE of the method with varying $\alpha$ and $\beta$ in the presence of Rayleigh noise. The settings of the parameters are also shown in the corresponding figures. The influence surfaces of $\alpha$ and $\beta$ on the steady-state MSE under different noise are presented in Fig. \ref{KRGMEE_MSE_RAY_AB} and Fig. \ref{mixedagaussianalbe}. Table \ref{diff_noise_alpha_time} and Table \ref{diff_noise_beta_time} show the pattern of the algorithm's performance with different $\alpha$ and $\beta$ in different scenarios. From the simulation results, one can obtain that the proposed algorithm works well for values of alpha or beta in the range $0.1 \leqslant \alpha  \leqslant 1.5$ or $1 \leqslant \beta  \leqslant 4$ under the given scenarios.

\begin{table*}
\centering
\caption{The steady-state MSE of the QKRGMEE algorithm with different $L$.}\label{diff_noise_L}
\begin{tabular}{lllllllll}
\hline
               & $L=5$    & $L=10$   & $L=15$   & $L=20$   & $L=30$     & $L=50$     & $L=80$     & $L=100$    \\ \hline
Rayleigh       & -32.09 & -32.79 & -33.03 & -33.21 & -33.39 & -34.59 & -35.08 & -35.12 \\
Mixed-Gaussian & -20.30 & -24.46 & -24.43 & -24.94 & -25.13 & -25.63 & -26.13 & -26.23 \\
Gaussian       & -29.99 & -30.52 & -30.98 & -31.08 & -31.30  & -31.41 & -31.85 & -31.98 \\
Mixed-noise    & -19.36 & -21.13 & -21.32 & -21.42 & -21.44 & -21.45 & -21.65 & -21.69 \\ \hline
\end{tabular}
\end{table*}

\begin{table*}
\centering
\caption{The steady-state MSE of the QKRGMEE algorithm with different $\gamma$.}\label{diff_noise_gamma_time}
\begin{tabular}{lllllllllllll}
\hline
                & \multicolumn{3}{c}{Rayleigh}                                                 & \multicolumn{3}{c}{mixed-Gaussian}                                           & \multicolumn{3}{c}{Gaussian}                                                 & \multicolumn{3}{c}{mixed-noise}                                              \\
                & \multicolumn{1}{c}{MSE} & \multicolumn{1}{c}{Time} & \multicolumn{1}{c}{H} & \multicolumn{1}{c}{MSE} & \multicolumn{1}{c}{Time} & \multicolumn{1}{c}{H} & \multicolumn{1}{c}{MSE} & \multicolumn{1}{c}{Time} & \multicolumn{1}{c}{H} & \multicolumn{1}{c}{MSE} & \multicolumn{1}{c}{Time} & \multicolumn{1}{c}{H} \\ \hline
KRLS             & -25.35                  & 0.00848                     & N/A                     & -7.24                   & 0.00848                     & N/A                     & -25.17                  & 8.63                     & N/A                     & -16.43                  & 0.00872                     & N/A                     \\
KRGMEE          & -29.65                  & 0.01197                    & N/A                     & -24.19                  & 0.01252                    & N/A                     & -30.94                  & 0.01292                    & N/A                     & -20.27                  & 0.01235                    & N/A                     \\
QKRGMEE($\gamma=0.01$) & -29.50                  & 0.01189                    & 10.65                   & -24.07                  & 0.01239                    & 10.21                   & -30.54                  & 0.01259                    & 5.59                    & -20.12                  & 0.01225                    & 21.9                    \\
QKRGMEE($\gamma=0.04$) & -29.01                  & 0.01176                    & 4.792                   & -23.71                  & 0.01226                    & 6.48                    & -30.50                  & 0.01228                    & 2.68                    & -19.45                  & 0.01216                    & 10.59                   \\
QKRGMEE($\gamma=0.1$)  & -28.91                  & 0.01170                    & 1.866                   & -22.86                  & 0.01218                    & 4.55                    & -30.47                  & 0.01201                    & 1.97                    & -19.33                  & 0.01213                    & 6.26                    \\
QKRGMEE($\gamma=0.15$) & -28.69                  & 0.01169                    & 1.436                   & -22.13                  & 0.01210                    & 4.26                    & -30.41                  & 0.01196                    & 1.579                   & -19.11                  & 0.01212                    & 4.73                    \\
QKRGMEE($\gamma=0.4$)  & -28.48                  & 0.01168                    & 1.015                   & -21.01                  & 0.01203                    & 3.92                    & -30.22                  & 0.01189                    & 1.201                   & -18.65                  & 0.01210                    & 2.571                   \\ \hline
\end{tabular}
\end{table*}

\begin{table*}
\centering
\caption{The steady-state MSE of QKRGMEE algorithm with different $\alpha$.}\label{diff_noise_alpha_time}
\begin{tabular}{llllllll}
\hline
               & $\alpha=0.1$  & $\alpha=0.4$    & $\alpha=0.8$    & $\alpha=1.0$      & $\alpha=2.0$      & $\alpha=4.0$      & $\alpha=8.0$  \\ \hline
Rayleigh       & -10.77 & -18.24 & -18.86 & -19.25 & -18.46 & -7.35  & -15.07    \\
Mixed-Gaussian & -11.16 & -24.99 & -24.99 & -24.82 & -20.75 & -8.06  & -0.97     \\
Gaussian       & -30.51 & -30.49 & -29.04 & -28.69 & -25.24 & -17.88 & -12.10    \\
Mixed-noise    & -10.77 & -18.24 & -18.86 & -19.25 & -18.46 & -7.35  & -0.85     \\ \hline
\end{tabular}
\end{table*}

\begin{table*}
\centering
\caption{The steady-state MSE of the QKRGMEE algorithm with different $\beta$.}\label{diff_noise_beta_time}
\begin{tabular}{llllllll}
\hline
               & $\beta=0.1$ & $\beta=0.4$    & $\beta=2.0$      & $\beta=4.0$      & $\beta=8.0$      & $\beta=15.0$     & $\beta=30.0$     \\ \hline
Rayleigh       & -27.33   & -28.56 & -30.26 & -30.63 & -31.15 & -32.07 & -32.89 \\
Mixed-Gaussian & -23.93   & -24.21 & -24.76 & -25.28 & -25.52 & -25.94 & -26.53 \\
Gaussian       & -26.69   & -27.50 & -28.28 & -29.16 & -30.06 & -30.13 & -30.19 \\
Mixed-noise    & -17.21   & -17.42 & -17.50 & -17.70 & -17.86 & -17.99 & -18.05 \\ \hline
\end{tabular}
\end{table*}

\subsection{EEG data processing}
\begin{figure*}[htbp] 
\centering  
\subfigure[Shape parameter ${\alpha}$.]{
\includegraphics[width=0.46\textwidth]{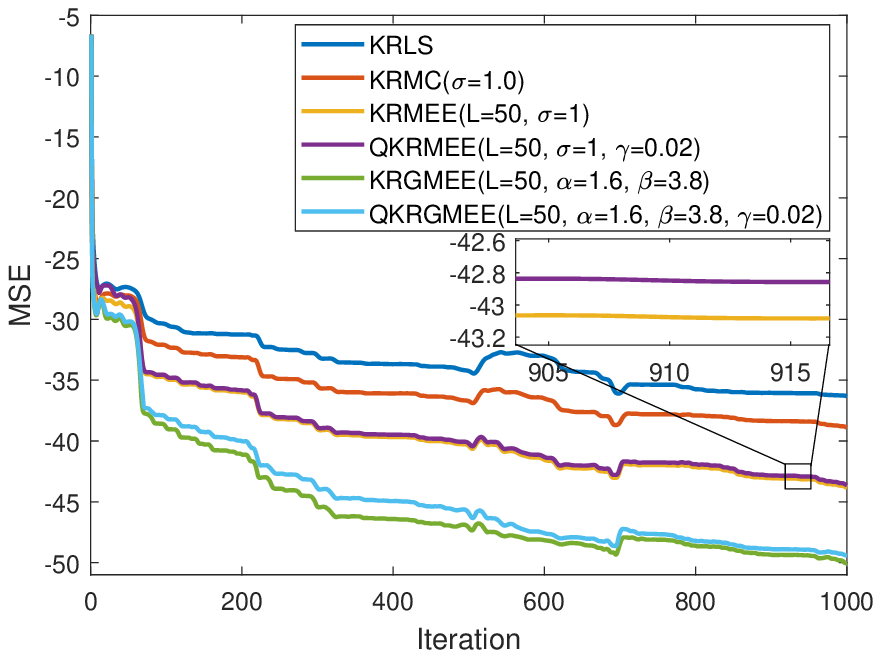}\label{EEG_MSE}
}
\quad
\subfigure[The length of the Parzen's window.]{
\includegraphics[width=0.46\textwidth]{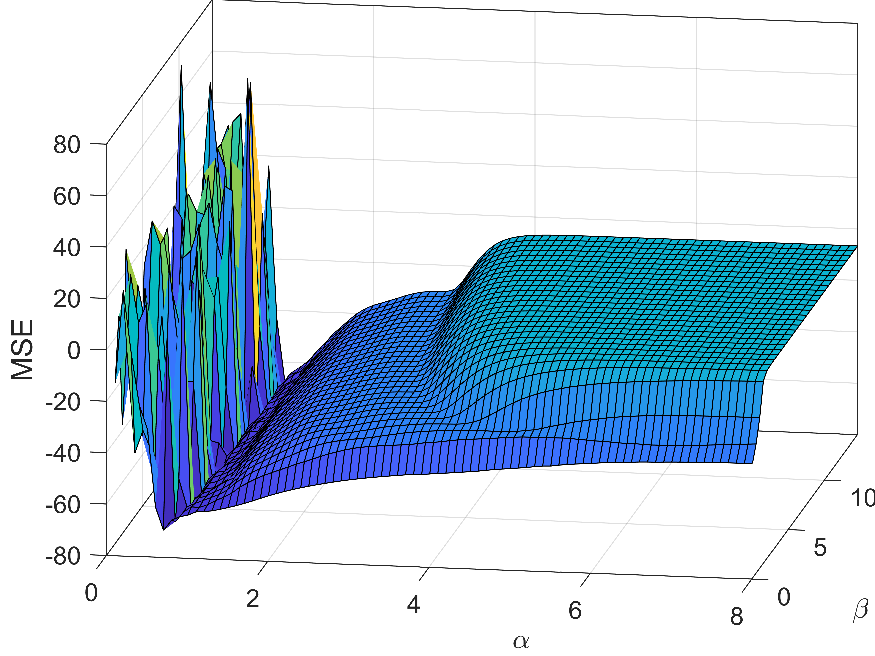}\label{EEG_MSE_AB}
}
\caption{The performance of the proposed algorithms for processing EEG data.}\label{EEG_MSE_AB_ALL}
\end{figure*}

In this part, we use our proposed QKRMEE and QKRGMEE algorithms to handle the real-world EEG data. By putting 64 Ag/AgCl electrodes and the expanded 10-20 system, the EEG data can be obtained from \cite{si2019different}. Moreover, the brain data is captured at a sampling rate of 500 Hz. The settings are presented in Fig. \ref{EEG_MSE}, and the results are displayed in Fig. \ref{EEG_MSE_AB_ALL}. Here, we use a segment of the FP1 channel data as the input. Fig. \ref{EEG_MSE} displays the convergence curves of QKRGMEE and its competitor, and Fig. \ref{EEG_MSE_AB} presents the surface of MSE with different $\alpha$ and $\beta$. 

The mean value of $H$ is 7.25 when $L=50$ and $r=0.02$, which shows that the quantizer can significantly reduce the computational burden of the algorithm without any significant degradation in the performance of the algorithm. It can be concluded that the performance of the QKRGMEE algorithm is much higher than that of the KRMEE algorithm and is comparable to that of the KRGMEE algorithm, even if the QKRGMEE algorithm has a lower computational complexity.

\section{Conclusion}\label{conclusion}
In this paper, We further refined the properties of the QGMEE criterion. On this basis, this QGMEE criterion was combined with the KRLS algorithm, and two new KRLS-type algorithms were derived, called QKRMEE and QKRGMEE respectively. QKRMEE algorithm is a special case of the QKRGMEE algorithm in which $\alpha=2$. Moreover, the mean error behavior, mean square error behavior, and computational complexity of the proposed algorithms are studied. In addition, simulation and real experimental data are utilized to verify the feasibility of the proposed algorithms. 

\section{Acknowledgements}\label{Acknowledgements}
This study was founded by the National Natural Science Foundation of China with Grant 51975107 and Sichuan Science and Technology Major Project No. 2022ZDZX0039, No.2019ZDZX0020, and Sichuan Science and Technology Program No. 2022YFG0343.

\bibliographystyle{unsrt}
\bibliography{cas-refs}

\begin{IEEEbiography}
[{\includegraphics[width=1in,height=1.25in,clip,keepaspectratio]{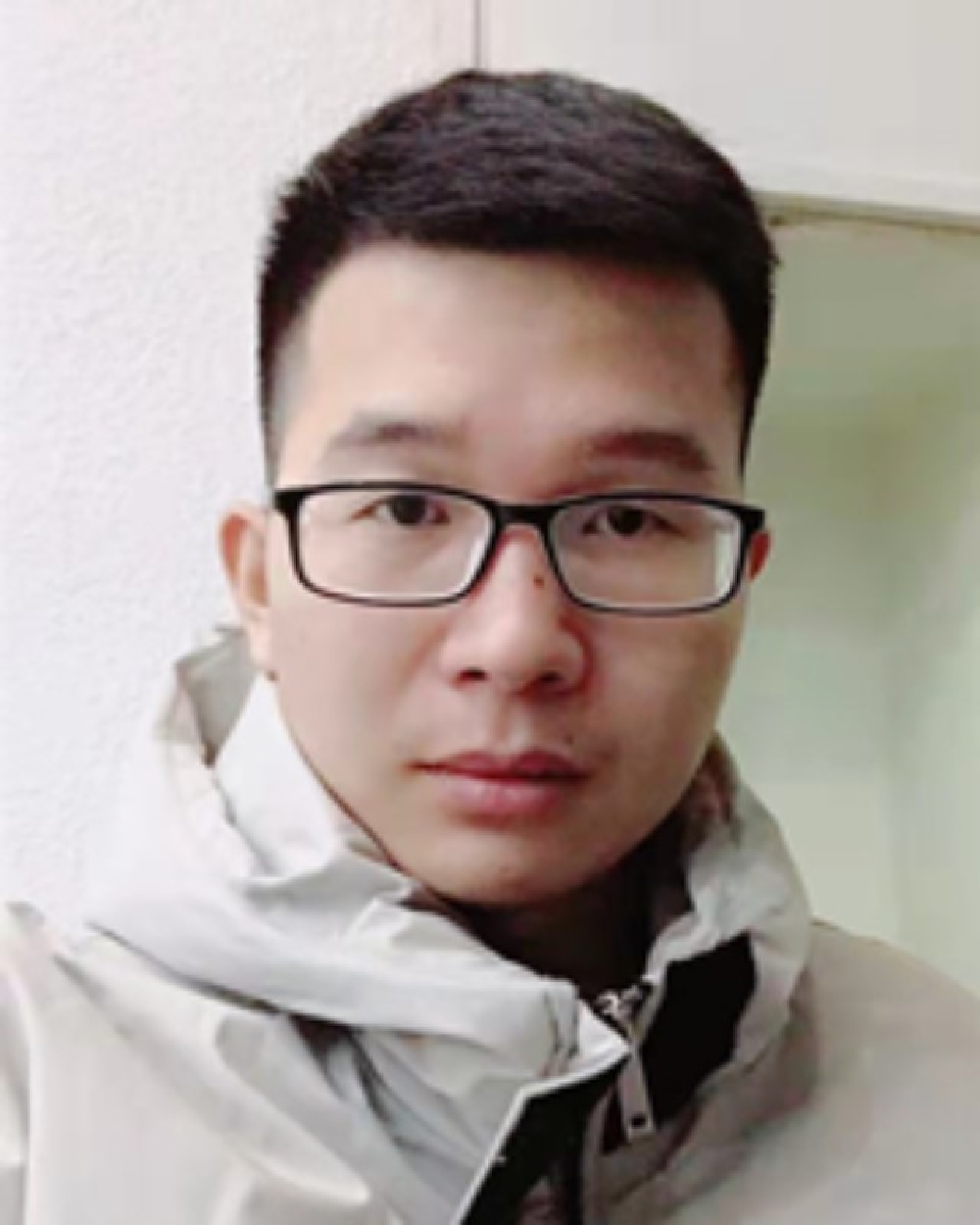}}] 
{Jiacheng He}{\space} received the B.S. degree in mechanical engineering from University of Electronic Science and Technology of China, Chengdu, China, in 2020. He is currently pursuing a Ph.D. degree in the School of Mechanical and Electrical Engineering, University of Electronic Science and Technology of China, Chengdu, China. His current research interests include information-theoretic learning, signal processing, and adaptive filtering.
\end{IEEEbiography}

\begin{IEEEbiography}
[{\includegraphics[width=1in,height=1.25in,clip,keepaspectratio]{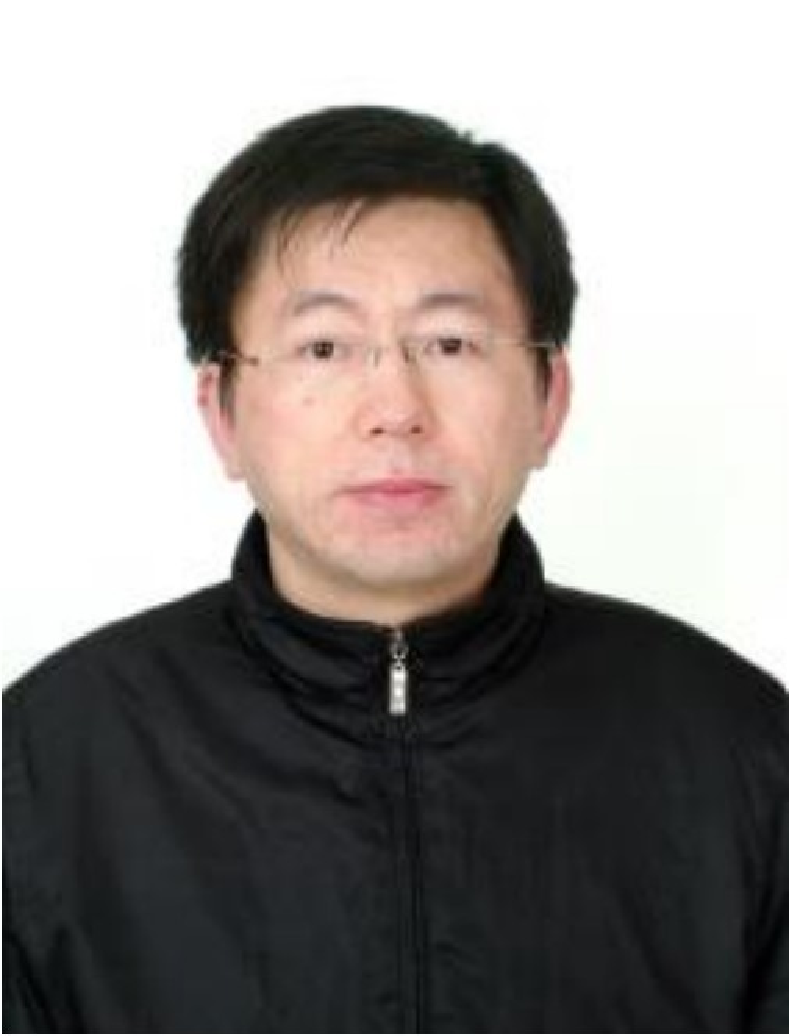}}] 
{Gang Wang}{\space} received the B.E. degree in Communication Engineering and the Ph.D. degree in Biomedical Engineering from University of Electronic Science and Technology of China, Chengdu, China, in 1999 and 2008, respectively. In 2009, he joined the School of Information and Communication Engineering, University of Electronic Science and Technology of China, China, where he is currently an Associate Professor. His current research interests include signal processing and intelligent systems.
\end{IEEEbiography}

\begin{IEEEbiography}
[{\includegraphics[width=1in,height=1.25in,clip,keepaspectratio]{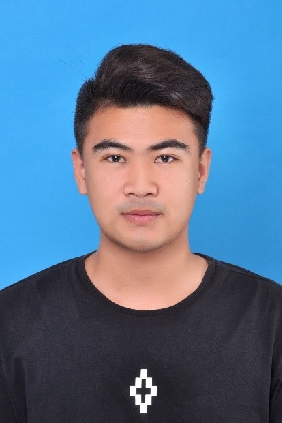}}] 
{Kun Zhang}{\space} received the B.S. degree in electronic and information engineering from Hainan University, Haikou, China, in 2018. He is currently working toward the Ph.D. degree in mechanical engineering with the Mechanical Engineering of the University of Electronic Science and Technology of China, Chengdu, China. His research interests include intelligent manufacturing systems, robotics, and its applications.
\end{IEEEbiography}

\begin{IEEEbiography}
[{\includegraphics[width=1in,height=1.25in,clip,keepaspectratio]{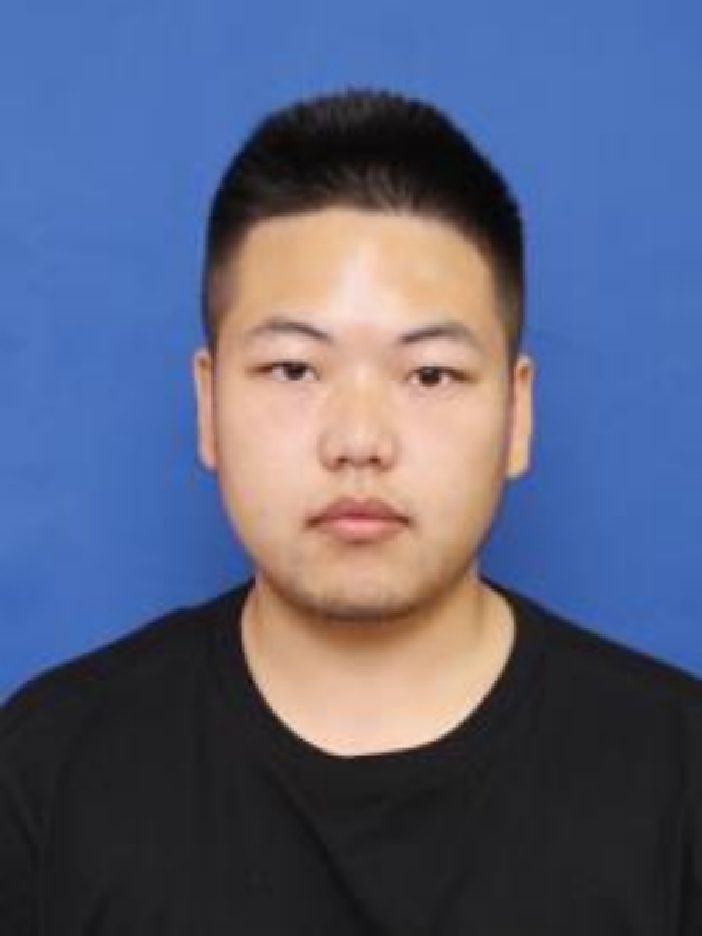}}] 
{Shan Zhong}{\space} received the B.E. degree in electrical engineering and automation with University of Technology, Chengdu, China in 2020. He is currently pursuing the M.S. degree in B.E. degree in communication engineering with School of Information and Communication Engineering, University of Electronic Science and Technology of China, Chengdu, China. His current research interests include signal processing and target tracking.
\end{IEEEbiography}

\begin{IEEEbiography}
[{\includegraphics[width=1in,height=1.25in,clip,keepaspectratio]{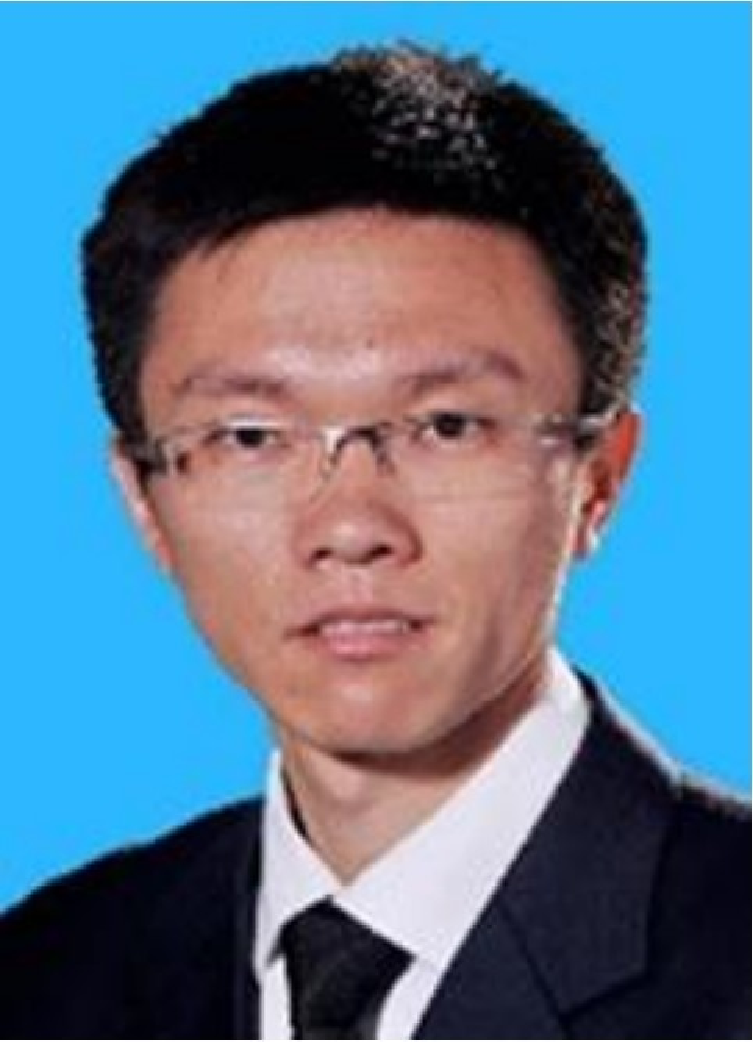}}] 
{Bei Peng}{\space} received the B.S. degree in mechanical engineering from Beihang University, Beijing, China, in 1999, and the M.S. and Ph.D. degrees in mechanical engineering from Northwestern University, Evanston, IL, USA, in 2003 and 2008, respectively. He is currently a Full Professor of Mechanical Engineering with the University of Electronic Science and Technology of China, Chengdu, China. He holds 30 authorized patents. He has served as a PI or a CoPI for more than ten research projects, including the National Science Foundation of China. His research interests mainly include intelligent manufacturing systems, robotics, and its applications.
\end{IEEEbiography}
\end{document}